\documentclass{jfm}
\usepackage{graphicx}
\usepackage{epstopdf, epsfig}
\usepackage{xcolor}
\usepackage{soul}
\usepackage{subcaption}   
\usepackage{amsmath}
\usepackage{bm}

\definecolor{cinnamon}{rgb}{0.82, 0.41, 0.12}

\shorttitle{Effect of coalescence in emulsions rheology}
\shortauthor{F. De Vita, M. E. Rosti, S. Caserta and L. Brandt}

\title[Rheology of emulsions]{On the effect of coalescence \\ on the rheology of emulsions}

\author{Francesco De Vita\aff{1}
  \corresp{\email{fdv@mech.kth.se}},
  Marco Edoardo Rosti\aff{1},
  Sergio Caserta\aff{2}
  \and Luca Brandt\aff{1}}

\affiliation{\aff{1} Linn\'e Flow Center and SeRC (Swedish e-Science Reseach Center), \\ KTH Mechanics, S-100 44 Stockhom, Sweden
  \aff{2} University of Naples "Federico II", Department of Chemical, Materials and Industrial Production Engineering, p.le Tecchio 80, 80125 Napoli, Italy}

\begin{document}

\maketitle

\begin{abstract}
  We present a numerical study of the rheology of a two-fluid emulsion in dilute and semidilute conditions.
  The analysis is performed for different capillary numbers, volume fraction and viscosity ratio under the assumption of negligible inertia and zero buoyancy force. The effective viscosity of the system increases for low values of the volume fraction and decreases for higher values, with a maximum for about 20\% concentration of the disperse phase. When the dispersed fluid has lower viscosity, the normalised effective viscosity becomes smaller than 1 for high enough volume fractions. To single out the effect of droplet coalescence on the rheology of the emulsion we introduce an Eulerian force which prevents merging, effectively modelling the presence of surfactants in the system. When the coalescence is inhibited the effective viscosity is always greater than 1 and the curvature of the function representing the emulsion effective viscosity vs. the volume fraction becomes positive, resembling the behaviour of suspensions of deformable particles. The reduction of the effective viscosity in the presence of coalescence is associated to the reduction of the total surface of the disperse phase when the droplets merge, which leads to a reduction of the interface tension contribution to the total shear stress. The probability density function of the flow topology parameter shows that the flow is mostly a shear flow in the matrix phase, with regions of extensional flow when the coalescence is prohibited. The flow in the disperse phase, instead, always shows rotational components. The first normal stress difference is positive, except for the smallest viscosity ratio considered, whereas the second normal difference is negative, with their ratio being constant with the volume fraction. Our results clearly show that the coalescence efficiency strongly affects the system rheology and neglecting droplet merging can lead to erroneous predictions.  
\end{abstract}

\section{Introduction}
Emulsions are a biphasic liquid-liquid system in which the two fluids are partially or totally immiscible. They are present in many biological and industrial applications such as waste treatment, oil recovery and pharmaceutical manufacturing. Emulsions are relevant also in the field of colloidal science where the accuracy and the control of the production process of functional materials rely on the knowledge of the complex microstructure of the suspension \citep{Xia2000}.

The study of the rheology of suspensions can be traced back to the pioneering work of \cite{Einstein1906,Einstein1911} who found that the effective viscosity $\mu_e$ of a dilute suspension of rigid sphere increases linearly with the volume fraction $\phi$ as $\mu_e = \mu_m(1+5\phi/2)$, where $\mu_m$ is the viscosity of the matrix phase. \cite{Batchelor1972} extended this relation including a term $\mathcal{O}(\phi^2)$, obtaining $\mu_e = \mu_m(1+5\phi/2+5\phi^2)$. These two analytical expression can well predict the suspension effective viscosity only in the dilute regime, at higher volume fraction no closed theory exist and we still rely on empirical relations. \cite{Eilers1941} proposed the following expression for emulsions $\mu_e = \mu_m [1+(5\phi/4)/(1-\phi/\phi_c)]$, which gives good approximation of the effective viscosity for low and high volume fractions. Here $\phi_c$ represent the maximum packing volume fraction, which depends on the properties of the suspended phase e.g. shape and deformability. \cite{Taylor1932} was the first to account for the deformation of the particles by introducing the viscosity ratio $\lambda$, defined as the disperse phase over the matrix phase viscosity, into the Einstein formula $\mu_e = \mu_m[1 + 2.5\phi(\lambda+0.4)/(\lambda+1)]$. Using perturbation analysis, \cite{Schowalter1968} where the first to find that for a suspension of deformable particles the first normal stress difference $N_1$ is positive and the second normal stress difference $N_2$ is negative, and later \cite{Choi1975} introduced a correction with $\mathcal{O}(\phi^2)$. \cite{Pal2002,Pal2003} derived expressions for the effective viscosity of infinitely dilute and concentrated emulsions using the effective medium approach. In general, all these previous relations are limited to the dilute regime for inertialess flows, are based on pair interaction between particles,  and typically assume limiting cases to model surface tensions effects (either going to zero or infinite). To overcome these limitations numerical simulations, providing access to all details of velocity and stresses in the system, play therefore an important role.

As concerns suspensions of rigid particles, many numerical investigations have been performed in the past years. For these systems the only relevant parameters are the solid volume fraction and the particle Reynolds number. \cite{Kulkarni2008} studied non-Brownian suspensions at finite particle Reynolds number, up to 5, using a Lattice Bolzman method. These authors found that inertia increases the particle contribution to the effective viscosity and breaks the fore-aft symmetry of the pair distribution function, see also \cite{Picano2013} for an analysis of the relative particle motion in the presence of inertia. \cite{Mari2014} reported shear thickening behaviour when a suspension of particles changes from the contactless to contact dominated regime and relate this effect to jamming. For a detailed review on suspensions of rigid particles, the reader is refereed to the recent perspective by \cite{Guazzelli2018}.

When considering deformable particles, emulsions being one relevant example, the scenario is further complicated by the occurrence of deformation, coalescence and breakup. \cite{Zhou1993} simulated numerically two-dimensional emulsions employing a boundary element method (BEM), which later has been extended to three-dimensional flows by \cite{Loewenberg1996}. \cite{Srivastava2016} investigated the inertial effects on emulsions using a front tracking method. These authors reported an inversion of the normal stress difference for increasing Reynolds number. \cite{Matsunaga2015}, also using a BEM method, investigated the rheology of a suspension of capsules up to 40\% of volume fraction and found that corrections order $\mathcal{O}(\phi^3)$ on the effective viscosity are negligible with respect to the rigid spheres. \cite{Rosti2018a} and \cite{Rosti2018} simulated suspensions of deformable particles described as a viscous hyperelastic solid material and showed that the effective viscosity of the suspension can be estimated by the Eilers formula, valid for rigid particles, when computing an effective volume fraction based on the mean deformation of the particles. As mentioned before, all these studies do not account for coalescence or breakup, which however plays a key role for the rheological behaviour of emulsions. The aim of this works is therefore to show how coalescence affects the rheology of emulsions.

The macroscopic properties of emulsions in shear flow strongly depend on their microstructure, mainly droplets size and distribution. The dynamics of the disperse phase is dictated by the interface deformation and the collision rate. In a dilute emulsion the interface deforms only under the action of the viscous stress exerted by the external flow. The relative importance  between this and the interfacial stress, which tends to keep the drop spherical, is known as the capillary number (Ca).  When Ca exceeds a critical value, droplets do not reach a steady state shape but break into two or more fragments in order to recover the balance between viscous and interfacial stresses. When the volume fraction increases, \emph{i.e.} in the non dilute regime, the flow can induce collisions between two or more drops. The outcome of a collision depends on the interaction force between the droplets. \cite{Chesters1991} proposed a conceptual framework to describe the complex dynamic of two colliding drops. The interaction can be thought of as the combined action of an external flow, responsible of the frequency, force and duration of the collision, and an internal flow which accounts for the deformation of the interface and the drainage of the thin liquid film between the two particles. When the collision duration is larger than the drainage time, droplets coalesce whereas in the opposite case they repel. In the first case the emulsion is attractive, in the latter case it is repulsive. Many experimental studies have been conducted to describe size evolution and deformation of droplets in microconfined shear flow \citep{Sibillo2006,Guido1998,Vananroye2006}. Interestingly, \cite{Caserta2012} have shown that the rheological curve, effective viscosity vs volume fraction, exhibits negative curvature for emulsions, contrary to the behaviour of particles (indeed the term $\mathcal{O}(\phi^2)$ in the Eilers formula is positive). In this work the authors reported also a phase separation for viscosity ratios smaller than one in large domains: they observed droplets rich and droplets poor regions oriented in the flow direction and alternated in the vorticity direction (vorticity banding). These results suggest that coalescence plays a key role in the rheological behaviour of emulsions and can affect their dynamics under shear flow.

Performing numerical simulations of emulsions in shear flow is a challenging problem due to the large scale separation between the external flow and the internal flow. The gap between the plates can be order of $\mu m$ while the thin liquid film where the Van der Waals force acts is order of $nm$ \citep{Chesters1991}. For this reason most of the numerical works on emulsions in literature neglect coalescence. Only few works that include coalescence are available: \cite{Shardt2013} investigated collision between two droplets in dilute emulsions using a Lattice Boltzman Method (LBM); \cite{Rosti2018b} present a numerical model for simulations of emulsions at moderate concentrations using a Volume of Fluid (VoF) approach and present results for different capillary numbers. Additionally, the presence of surfactants (surface active agents) can significantly affect the interface dynamics in terms of droplets deformability and distribution evolution. The modelling of this mechanisms in numerical simulation is not trivial \cite{Khatri2011,Soligo2019}. An alternative to previous mentioned methods is to employ mesoscale methods which consider also the thermal effects, as for example in \cite{Sega2013}. In this study, we complete the study on the rheology of emulsions by including the effect of the viscosity ratio and, more important, the role of coalescence/breakup on the macroscopic behavior. The use of a Volume of Fluid technique allows to take into account the coalescence of droplets, while we model repulsive short-range interactions.

The paper is structured as follow: in section \ref{sec:numerics} we describe the governing equations, the numerical method and the setup we adopt; in section \ref{sec:collision} we present a fully Eulerian framework to introduce a collision force in the solver which is able to delay or prevent the coalescence between droplets; in section \ref{sec:results} we discuss the results in term of effective viscosity, stresses budget, and normal stresses for emulsions with different  coalescence rates. Finally we summarize the main findings and draw some final conclusions in section \ref{sec:conclusion}.

\section{Numerical method and setup}\label{sec:numerics}

In this study we investigate the rheology of emulsions in shear flow simulating a liquid-liquid biphasic system in a simple Couette flow. The flow dynamics is governed by the incompressible Navier-Stokes equations which in the one-fluid formulation for multiphase flows read 
\begin{subequations}
  \begin{align}
    \label{eqn:navier-stokes}
    \frac{\partial u_i}{\partial x_i} &= 0, \\
    \rho\left(\frac{\partial u_i}{\partial t} + u_j\frac{\partial u_i}{\partial x_j}\right) &= -\frac{\partial p}{\partial x_i} + \frac{\partial}{\partial x_j}\left(2\mu D_{ij}\right) + \sigma \kappa \delta_s n_i.
  \end{align}
\end{subequations}
Here $u_i$, with $i = 1,2,3$, are the velocity components in the three cartesian coordinates $x_1, x_2$ and $x_3$, $p$ the pressure field, $\rho$ and $\mu$ the local density and viscosity, $\mathbf{D}$ the rate of deformation tensor $D_{ij} = \left(\partial u_i / \partial x_j + \partial u_j / \partial x_i\right)/2$, $\sigma$ the interfacial tension coefficient, $\kappa$ the curvature of the interface, $n_i$ the $i-th$ component of the unit normal vector $\mathbf{n}$ to the interface and $\delta_s$ the dirac function which express that the interfacial tension force acts only at the interface between the two fluids. 

To track in time the position of the interface we employ a VoF technique based on the multi-dimensional tangent of hyperbola interface capturing (MTHINC) method \citep{Ii2012}. To identify the two fluids we define a color function $\mathcal{H}(\mathbf{x},t)$ so that $\mathcal{H} = 1$ in one fluid and $\mathcal{H} = 0$ in the other. The Volume of Fluid function $\mathcal{T}(\mathbf{x},t)$ is defined as the cell-average value of the color function
\begin{equation}
  \mathcal{T}(\mathbf{x},t) = \frac{1}{\delta \mathcal{V}}\int_{\delta \mathcal{V}} \mathcal{H}(\mathbf{x}',t)\,d\mathcal{V}'
\end{equation}
and represents the volume fraction in each cell of the domain. The VoF function is advected by the flow field as
\begin{equation}
  \label{eqn:vof}
  \frac{\partial \mathcal{T}}{\partial t} + \frac{\partial u_j \mathcal{H}}{\partial x_j} = \mathcal{T}\frac{\partial u_j}{\partial x_j}.
\end{equation}
The key point of the MTHINC method is to approximate the color function with a tangent of hyperbola
\begin{equation}
  \mathcal{H}(X,Y,Z) \approx \hat{\mathcal{H}}(X,Y,Z) = \frac{1}{2}(1 + \text{tanh}(\beta(P(X,Y,Z)+d)))
\end{equation}
where $(X,Y,Z) \in [0,1]$ is a local coordinate system, $\beta$ a sharpness parameter, $d$ a normalization factor and $P$ a three-dimensional surface function which can be either linear (plane) or quadratic (curved surface) with no additional cost. This discretization allows to solve the fluxes of equation \eqref{eqn:vof} by integration of the approximated color function in each computational cell. The material properties of the two fluids are linked to the VoF function $\mathcal{T}$ as follow
\begin{equation}
  \begin{aligned}
    \rho(\mathbf{x},t) = \rho_1\mathcal{T}(\mathbf{x},t) + \rho_0(1-\mathcal{T}(\mathbf{x},t)) \\
    \mu(\mathbf{x},t) = \mu_1\mathcal{T}(\mathbf{x},t) + \mu_0(1-\mathcal{T}(\mathbf{x},t))
  \end{aligned}
\end{equation}
where the subscript 1 stands for the disperse phase, the subscript 0 for the carrier fluid and $\mathcal{T}$ is equal to 1 in the disperse phase and 0 in the carrier fluid. Finally the surface tension force is approximated using the Continuum Surface Force (CSF) approach \citep{Brackbill1992}
\begin{equation}
  \sigma \kappa \delta_s n_i = \sigma \kappa \frac{\partial \mathcal{T}}{\partial x_i}.
\end{equation}
See \cite{Rosti2018b} for a detailed description and validation of the code employed in this work.

\begin{figure}
  \centering
  \includegraphics[width=0.8\textwidth]{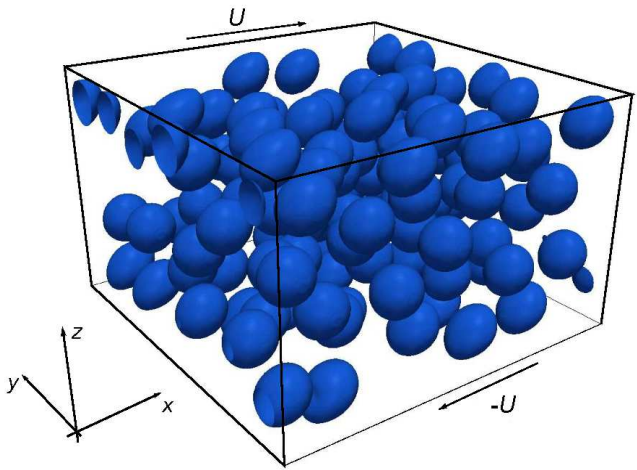}
  \caption{Sketch of the computational domain and coordinate system used.\label{fig:sketch}}
\end{figure}

\subsection{Flow configuration}
A random monodisperse distribution of droplets, with radius $r$, is initialized in a bi-periodic system ($x$ being the streamwise direction and $y$ the spanwise) delimited by two walls ($z$ direction) moving in opposite direction with velocity $U$ at a constant distance $L_z$. Choosing as reference length the initial radius of the droplets $r$, the size of the domain is 16$r$ x 16$r$ x 10$r$ ($x$, $y$ and $z$, respectively), as sketched in figure \ref{fig:sketch}. This configuration has been widely adopted in literature for the study of rheology of suspensions of rigid and deformable particles and emulsions  \citep[see][]{Picano2013,Rosti2018a,Rosti2018b}. The flow is governed by four non-dimensional parameters, namely the Reynolds number $Re$, the capillary number $Ca$, the viscosity ratio $\lambda$ and the volume fraction $\phi$, with
\begin{equation}
  Re = \frac{\rho_0 \dot{\gamma} r^2}{\mu_0}, \quad Ca = \frac{\mu_0 \dot{\gamma}r}{\sigma}, \quad \lambda = \frac{\mu_1}{\mu_0}.
\end{equation}
The applied shear rate is equal to $\dot{\gamma} = 2U/L_z$ and, in order to keep inertial effect negligible, is chosen to give Reynolds number equal to 0.1. The interface tension coefficient $\sigma$ is chosen so that the capillary number based on the initial radius varies between 0.05 and 0.2. The simulations are performed with fixed Reynolds number, for different viscosity ratios, $\lambda = [0.01, 0.1, 1]$ and for four volume fractions, $\phi = [0.00164, 0.1, 0.2, 0.3]$, the first one corresponding to one single droplet initially located at the center of the domain. In all the cases the density of the two fluid is the same, thus there are no buoyancy effects. All the simulations have been performed with a mesh of 256x256x160 cells, which correspond to a grid size of $\Delta = r / 16$. The independence of the results on the grid size, measured by the suspension effective viscosity and surface area evolution, was verified by performing simulations with half and double resolution.

\section{Collision force model}\label{sec:collision}

\begin{figure}
  \centering
  \includegraphics[width=0.7\textwidth]{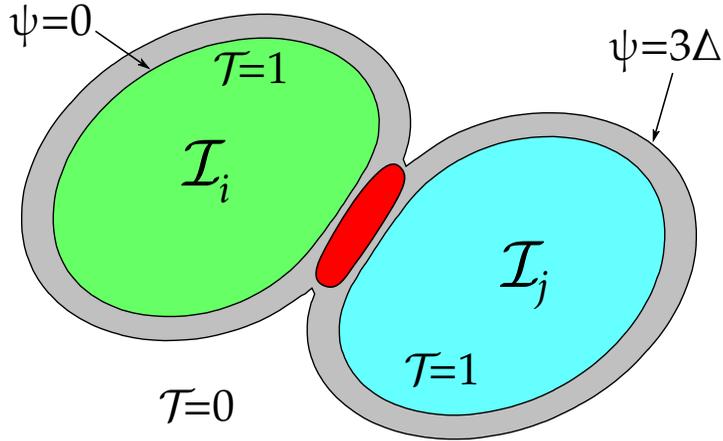}
  \caption{Sketch of the collision model: the gray area represents the region of propagation of the distance function $\psi$, the external contour corresponding to $\psi = 3\Delta$. The red region represent the area where the force is applied.\label{fig:collision_sketch}}
\end{figure}

The outcome of a collision between two droplets is function of hydrodynamic forces, capillary forces and geometric parameters, as radius and relative distance. Droplets coalesce when the capillary number is smaller than a critical value $Ca_c$ whereas they slide away when the capillary number is larger than $Ca_c$ \citep{Shardt2013}. To investigate the effect of coalescence on the rheology of emulsions we introduce a repulsive force with the aim of completely inhibiting the merging of droplets. From a physical point of view, this force can be seen as a model for the presence of surfactants in the emulsion, which alter the value of the interfacial tension coefficient and change the capillary number. The action of surfactants occurs on nanoscales, which cannot be resolved on our mesh and therefore needs to be modeled. To prevent coalescence one should impose a very strong interfacial tension coefficient which would make simulations expensive by requiring a very small time step. Because we are only interested in inhibiting coalescence, we introduce an Eulerian repulsive force of the same form as a lubrication force\citep{Bolotnov2011}
\begin{equation}
  \label{eqn:force}
  \mathbf{F_c} = \mu_0 r U \left(\frac{a}{\psi} + \frac{b}{\psi^2}\right)\mathbf{n}
\end{equation}
where $a$ and $b$ are two coefficients and $\psi$ is the signed distance from the interface. Operationally, we reconstruct every time step the distance function from the interface (defined by the VoF function $\mathcal{T}$) as follows: first we initialize the distance function $\psi_0$ following \cite{Albadawi2013}
\begin{equation}
  \psi_0 = \left(2\mathcal{T}-1\right) 0.75 \Delta
\end{equation}
with $\Delta$ the grid spacing; then the distance function is propagated in a region of thickness $3 \Delta$ from the interface solving the redistancing equation
\begin{equation}
  \frac{\partial \psi}{\partial \tau} + S\left(\psi_0\right)\left(|\nabla\psi| - 1\right) = 0
\end{equation}
where $S\left(\psi_0\right)$ is the sign function and $\tau$ is an artificial time; the equation is marched in time with step $\Delta \tau = 0.1\Delta$. We solve the previous equation using the second order redistancing algorithm of \cite{Russo2000}. At the beginning of the simulation, every droplet is tagged with an integer number $\mathcal{I}$, going from 1 to the total number of droplets. This index is used to determine where to apply the collision force, as sketched in figure \ref{fig:collision_sketch}. 
The grey area in the figure represent the region, of thickness $3\Delta$, where the distance function is propagated and the red region is the area where the force is applied. This region is found by checking, inside the grey area, if in a stencil $7\Delta\times7\Delta\times7\Delta$ there are two different values of $\mathcal{I}$. If this condition is verified, the interaction force is computed following \eqref{eqn:force}. Two additional indexes are used to identify the walls and model collision with them, also to avoid mass losses at the boundary. Every time step, the droplet index $\mathcal{I}$ is updated based on the new position of $\mathcal{T}$: if $\mathcal{T} < 0.5$ we set $\mathcal{I} = 0$ otherwise we search in a stencil $3\Delta\times3\Delta\times3\Delta$ the non-zero value of $\mathcal{I}$. To avoid instabilities resulting from the application of the force directly on the surface, the force is applied only if $\psi$ is larger then $\sqrt{2}\Delta/ 2$ in 2D and  $\sqrt{3}\Delta/ 2$ in 3D. The  algorithm currently implemented to advance in time the index function does not handle topological changes, \emph{i.e.} break-up or coalescence, hence we need to ensure that none of the droplets coalesces or breaks during the simulations. By trial and error, we found that the minimum values of the coefficients in equation \eqref{eqn:force} that always prevent coalescence, for the given non-dimensional parameters of the flow, are $a = 55$ and $b = 3.5$. These values have been used to obtain the results presented in the next section.

\begin{figure}
  \centering
  \includegraphics[width=0.75\textwidth]{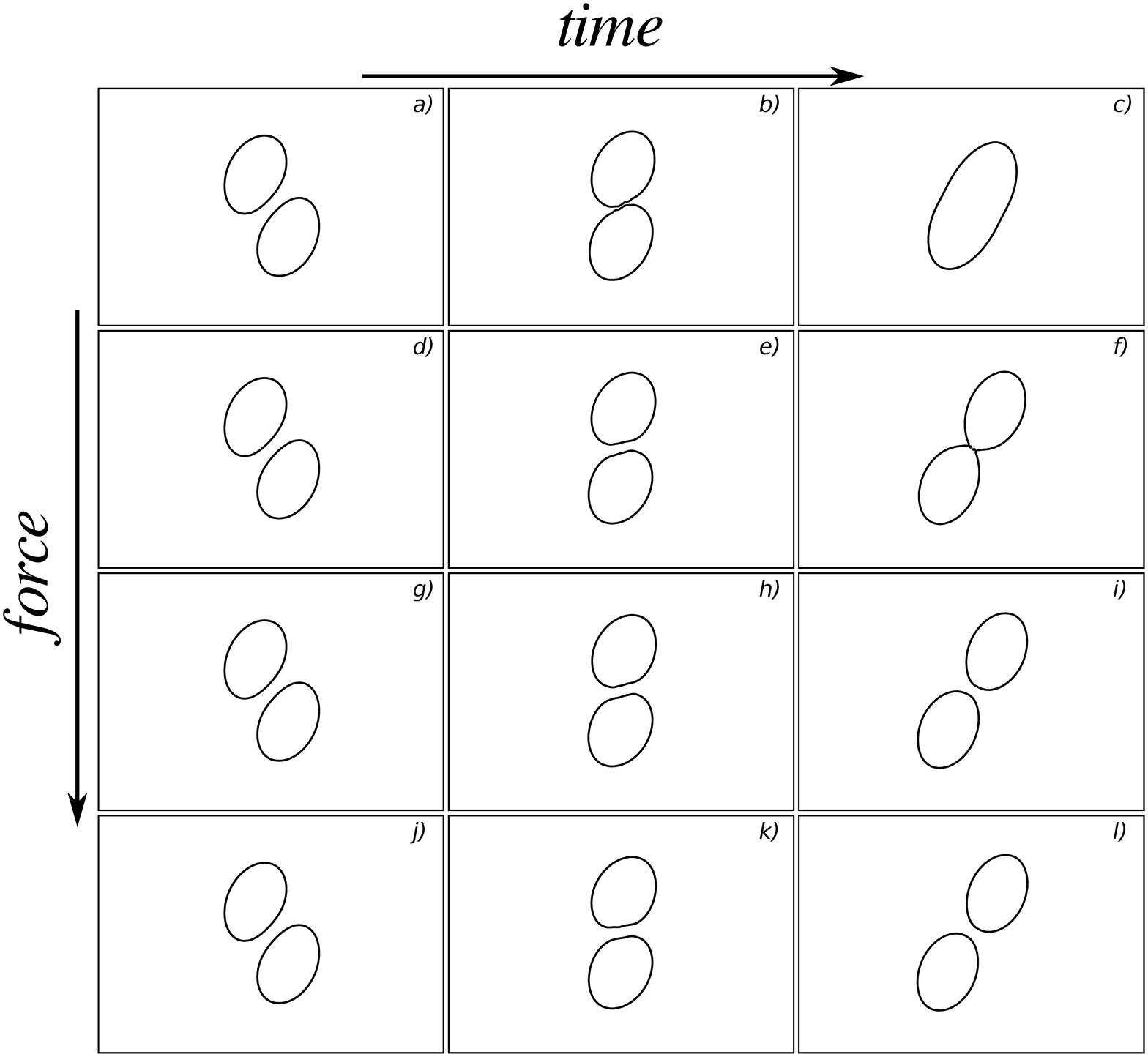}
  \caption{Time evolution of the interface at three different time instants; from left to right: $t = 10$, $t = 15$ and $t = 20$. Figures (a)-(b)-(c) zero force; figures (d)-(e)-(f) $a = 13.75$ and $b = 0.875$; figures (d)-(e)-(f) $a = 27.5$ and $b = 1.75$; figures (g)-(h)-(i) $a = 55$ and $b = 3.5$.\label{fig:2dinter}}
\end{figure}
Before moving to the next section, we show the effect of the repulsive force \eqref{eqn:force} on the interaction between two droplets in a simple shear flow and the sensibility to the coefficients $a$ and $b$. We consider two drops of equal initial radius $r = 1$, the length of the domain is $12r$ and the height is $6r$. The two droplets are placed in the center of the domain with a horizontal shift $\Delta x = 3$ (half of the channel height) and with a vertical shift $\Delta y = 1.5$. The particle Reynolds number $Re=0.1$, the capillary number $Ca= 0.1$ and the viscosity and density ratio equal to unity. Note that this setup is similar to that adopted in other numerical and experimental studies \citep{Guido1998,Shardt2013}. Figure \ref{fig:2dinter} shows the interface location at three different time instants, $t = 10$, $t = 15$ and $t = 20$, from the left to the right. Each row of the figure, from top to bottom, corresponds to a different case: first row (panels (a)-(b)-(c)) corresponds to the case of zero force, when droplets are free to coalesce (this is the case with maximum drainage velocity); second row (panels (d)-(e)-(f)) corresponds to the case with coefficients $a = 13.75$ and $b = 0.875$; third row (panels (g)-(h)-(i)) corresponds to the case with coefficients $a = 27.5$ and $b = 1.75$; last row (panels (j)-(k)-(l)) corresponds to the case with coefficients $a = 55$ and $b = 3.5$, same as in the results section. As soon as the droplets come closer to each other, they start to flatten and to deform. As a consequence of the applied shear and of the interaction of the particle with the local flow field they start to rotate. For the adopted capillary number and initial displacement and in absence of any collision force, the two drops coalesce at a certain instant, $t \approx 16$. If the applied force is small (second row) the time needed to actually merge is larger and the coalescence is delayed. This is an interesting result because by tuning the value of the force it is possible to change the timescale of the drainage of the film between the droplets. Finally, the bottom row shows the case of a fully repulsive force, completely preventing coalescence.

\begin{figure}
  \centering
  \includegraphics[width=0.8\textwidth]{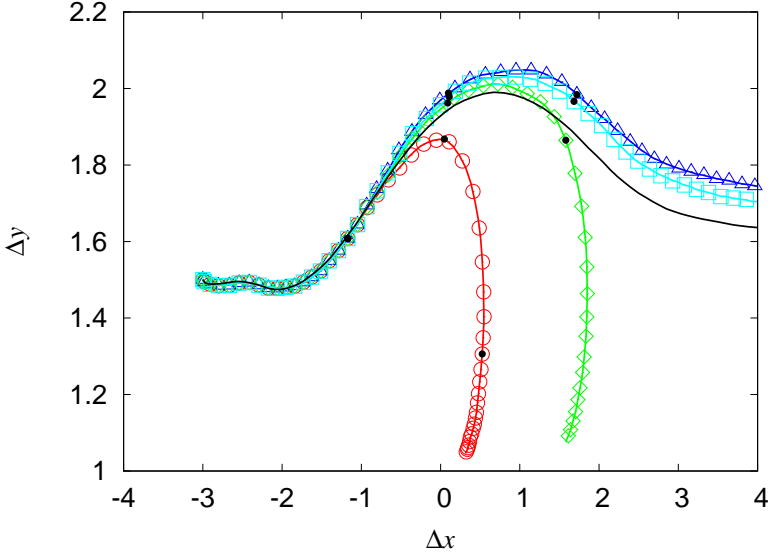}
  \caption{Plot of relative distance between droplets $\Delta y$ vs $\Delta x$: case with no force (\textcolor{red}{$\circ$}); case with $a = 13.75$ and $b = 0.875$ (\textcolor{green}{$\diamond$}); case with $a = 27.5$ and $b = 1.75$ (\textcolor{cyan}{$\Box$}); case with $a = 55$ and $b = 3.5$ (\textcolor{blue}{$\triangle$}). Black dots (\textbullet) correspond to the time instant in figure \ref{fig:2dinter}. The black line correspond to the simulation without collision force and with double resolution.\label{fig:shift}}
\end{figure}

Figure \ref{fig:shift} shows the time history (from the left to the right in a curvilinear system) of the vertical displacement $\Delta y$ over the horizontal displacement $\Delta x$ for the four cases of figure \ref{fig:2dinter}. When the droplets coalesce both the horizontal and vertical displacement decrease quickly, as illustrated by the red and green curves in figure \ref{fig:shift}. In the opposite case, after the collision, droplets continue to move away in the horizontal direction with a steady state value of the vertical displacement. Although the force in the fourth case is twice as large as that in the third case the final vertical displacement differs only of about 3\%, showing that the force does not affect significantly the dynamic of the droplets after the collision. To check the effect of the grid size we also performed one simulation without collision force and with double resolution, reported with a black line in the same figure. The final value of the vertical displacement for the cases with force is about 6\% different than the case with double resolution.

\section{Results}\label{sec:results}

We start our analysis by first considering the cases without the repulsive force given by equation \eqref{eqn:force}. Although the two fluid considered are both Newtonian, emulsions can exhibit non-Newtonian behaviour like shear thinning, normal stress differences or viscoelasticity \citep{Foudazi2015}. A first parameter used to characterize the rheological behaviour of a suspension is the effective shear viscosity of the system, $\mu_e$. The wall-normal gradient of the streamwise velocity gradient at the walls can be used to evaluate the effective viscosity of an emulsion as
\begin{equation}
  \label{eqn:effective}
  \mu_e = \frac{<\mu_0 \frac{\partial u}{\partial z}|_w>}{\dot{\gamma}}
\end{equation}
where the symbol $< >$ indicates that the quantity has been averaged in the two homogeneous directions and in time. Following \cite{Yang2016}, it can be shown that the wall normal velocity gradients at the wall is equivalent to the bulk shear stress. Here, we show how to include the interfacial tension contribution inside the shear stress balance. At steady state the streamwise momentum equation, after averaging in the homogeneous direction, reduces to
\begin{equation}
  0 = - \rho \frac{\partial <uw>}{\partial z} + \frac{\partial <2 \mu D_{13}>}{\partial z} + <f_1>
\end{equation}
where $f_1$ represent the streamwise component of the interface tension force. This equation can be rewritten as
\begin{equation}
  \label{eqn:shearbalance}
  0 = \frac{\partial}{\partial z}\left(-\rho <uw> + <2 \mu D_{13}> + \int_z <f_1>\,dz \right);
\end{equation}
where we define $\mathcal{G}(z) = \int_z <f_1>\,dz$ the integral of $<f_1>$ in the wall-normal direction. To compute this integral we approximate the relation $\partial \mathcal{G} / \partial z = <f_1>$ with a finite difference scheme and invert it. To determine the constant of integration we impose $\mathcal{G} = 0$ at the walls. We can now integrate equation \eqref{eqn:shearbalance} from the wall, $z = 0$, to a generic section $z=\zeta$
\begin{equation}
  \begin{aligned}
  -\rho <uw>|_{0} + &<2 \mu D_{13}>|_{0} + <\mathcal{G}>|_0 = \\
  &-\rho <uw>|_{\zeta} + <2 \mu D_{13}>|_{\zeta} + <\mathcal{G}>|_{\zeta}.
  \end{aligned}
\end{equation}
The first term on the left-hand-side is zero because velocity is zero at the wall and the last term on the left-hand-side is zero due to the imposed boundary condition, therefore
\begin{equation}
  \begin{aligned}
  <2 \mu D_{13}>|_{0} = -\rho <uw>|_{\zeta} + <2 \mu D_{13}>|_{\zeta} + <\mathcal{G}>|_{\zeta}.
  \end{aligned}
\end{equation}
Because of the arbitrary choice of the section $\zeta$, we can average the right-hand-side in the wall normal direction
\begin{equation}
    <2 \mu D_{13}>|_{0} = \frac{1}{L_z}\int_{0}^{L_z}\left(-\rho <uw>|_{\zeta} + <2 \mu D_{13}>|_{\zeta} + <\mathcal{G}>_{\zeta}\right)\,dz
\end{equation}
or similarly
\begin{equation}
  \label{eqn:stress-balance}
  <2 \mu D_{13}>|_{0} = - <<\rho uw>> + <<2 \mu D_{13}>> + <<\mathcal{G}>>
\end{equation}
where the symbol $<< >>$ indicates that the quantity has been averaged in the whole domain. Providing that the first term on the right-hand-side is negligible, we have proved that the wall-normal shear stress is a measure of the bulk shear stress in the domain, given by the sum of the viscous part and the interface tension contribution. Before proceeding to quantify the shear stress components we need to underline two aspects: \emph{i}) the interfacial term $\mathcal{G}$ at the wall is zero only if there are no droplets reaching the wall and this is always verified in our simulations, as shown in the following section; \emph{ii}) in order to consider our solution at steady state we average in time over several shear unit (typically between $20$ and 40) and check if the mean value of the shear rate varies less than few percent, therefore all the quantities have to be considered also as averaged in time.

\begin{figure}
  \centering
  \begin{subfigure}{0.45\textwidth}
    \includegraphics[width=\textwidth]{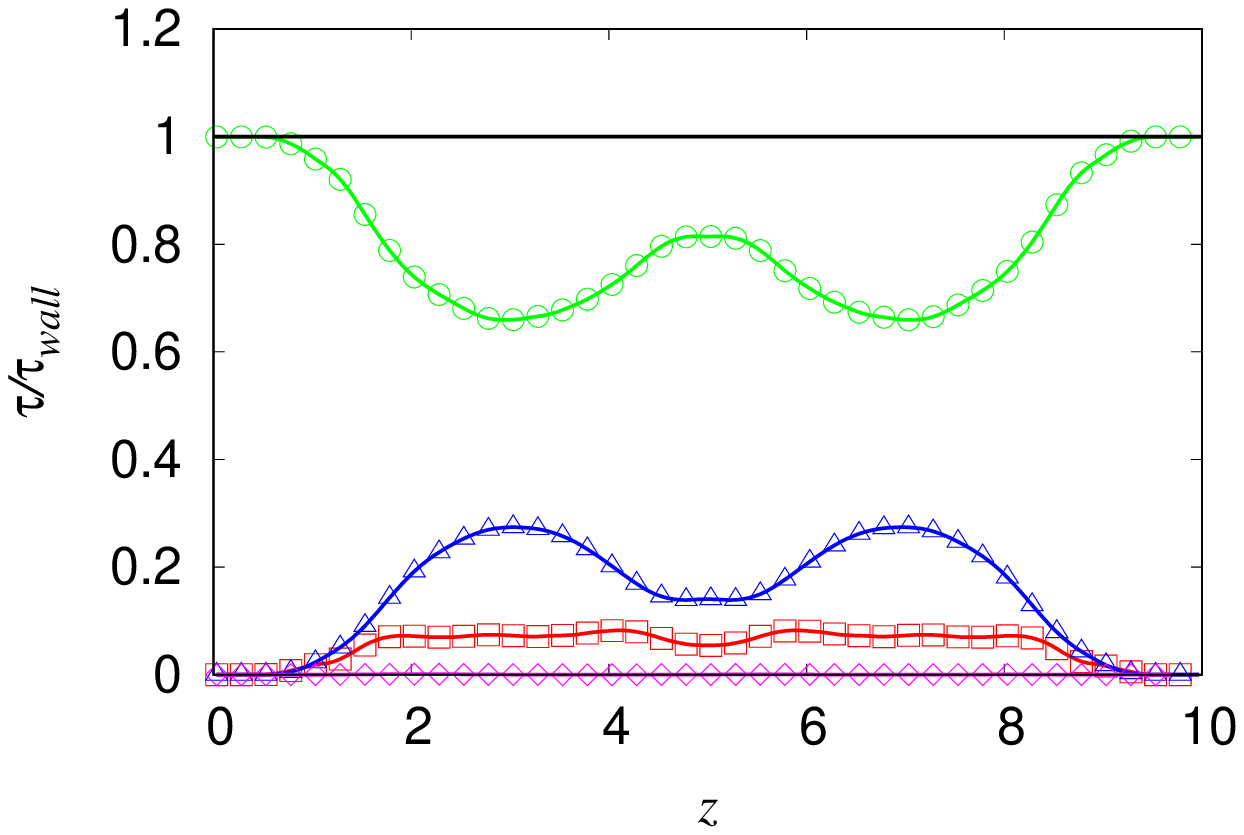}
  \end{subfigure}
  \begin{subfigure}{0.45\textwidth}
    \includegraphics[width=\textwidth]{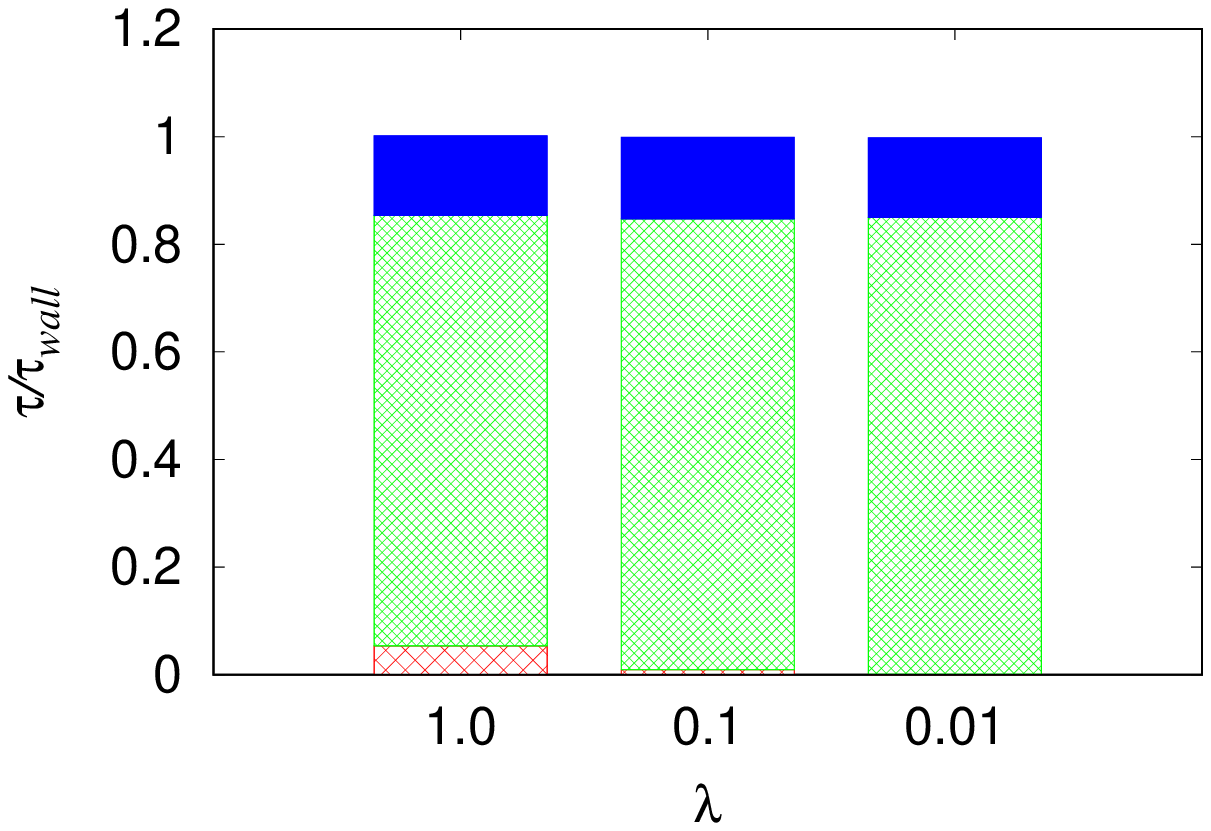}
  \end{subfigure}
  \caption{(Left panel) Stress distribution along the wall normal direction $z$ (i.e. the direction of the velocity gradient) for the case with $Ca = 0.1$, $\phi = 0.1$ and $\lambda = 1.0$: disperse phase viscous stress (\textcolor{red}{$\Box$}); outer fluid viscous stress (\textcolor{green}{$\circ$}); $\mathcal{G}$ (\textcolor{blue}{$\triangle$}); advection (\textcolor{magenta}{$\diamond$}). (Right panel) Histogram of the stress components for the case with $Ca = 0.1$, $\phi = 0.1$ and three viscosity ratio: interfacial force (solid blue); outer fluid viscous stress (green dense net); disperse phase viscous stress (red sparse net); The stresses are normalized with the wall shear stress.\label{fig:stress-components}}
\end{figure}

To illustrate the outcome and indications of such an analysis, we display in figure \ref{fig:stress-components} (left panel) the profile of the stress components over the wall normal direction $z$, that is the direction of the velocity gradient, obtained from equation \eqref{eqn:stress-balance}. As already anticipated, the advection term is small (order $10^{-6}$) and for this reason will be neglected in the discussion hereinafter. The main contribution to the total shear stress is given by the viscous stress of the outer fluid, which is also the only non-zero term at the walls, and the sum of all the components is constant along the vertical direction. This confirm that the shear stress at the wall is equal to the bulk shear stress and that the effective viscosity can be evaluated using the wall-normal gradient of the streamwise velocity at the wall, see equation \eqref{eqn:effective}. On the right panel of the same figure, we report the histogram of the shear stress components for three different values of the viscosity ratio. Here, we note that the contribution due to the viscous stress of the dispersed phase (red sparse net) is almost negligible for $\lambda$ lower than 1, while the interfacial tension (in blue) remains constant, being the volume fraction constant. 

\begin{figure}
  \centering
  \begin{subfigure}{0.49\textwidth}
    \includegraphics[width=\textwidth]{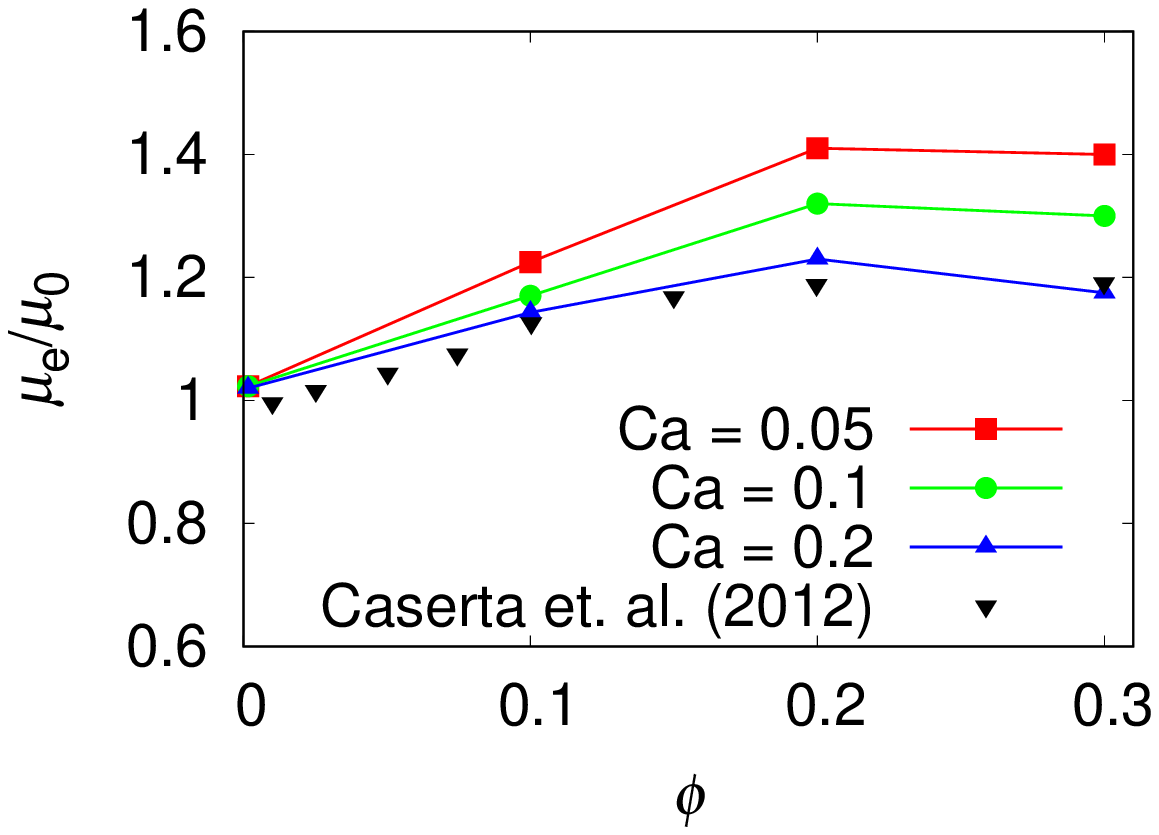}
  \end{subfigure}
  \begin{subfigure}{0.49\textwidth}
    \includegraphics[width=\textwidth]{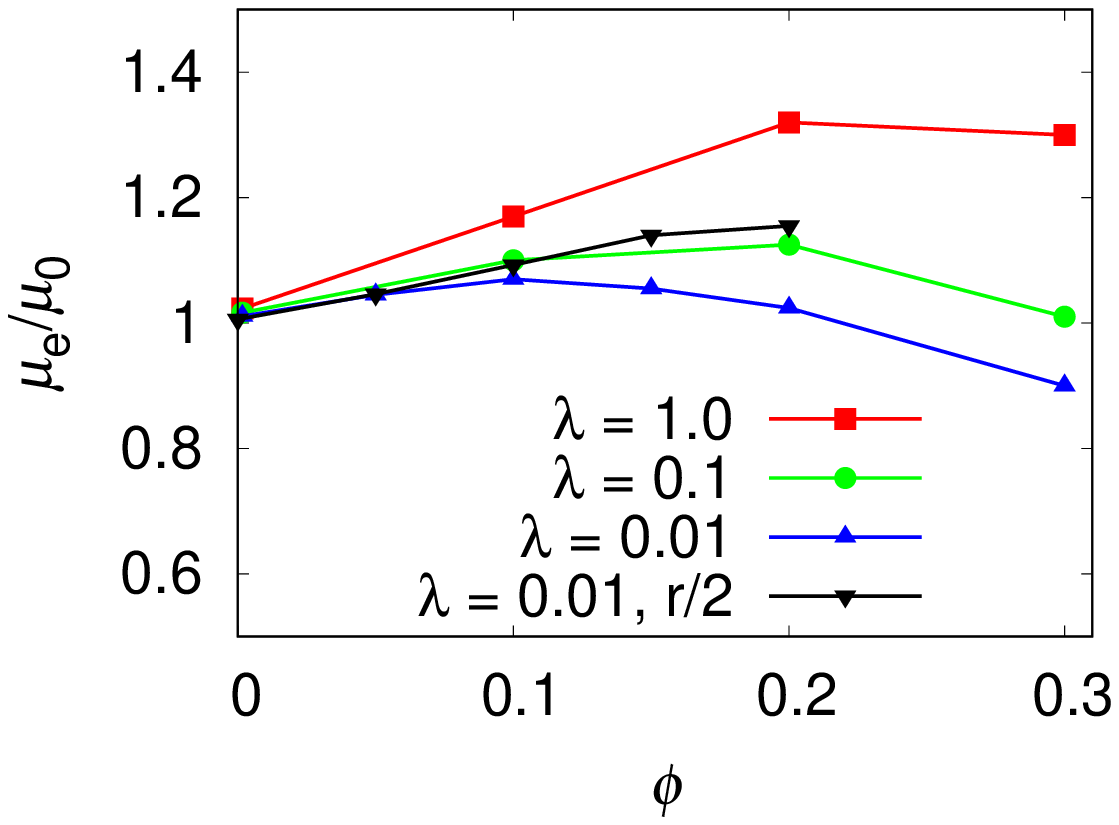}
  \end{subfigure}
  \caption{Effective viscosity $\mu_e$ as function of the volume fraction: $\lambda = 1.0$ and different $Ca$ (left panel); $Ca = 0.1$ and different $\lambda$ (right panel). The effective viscosity is normalized with the outer fluid viscosity.\label{fig:effective}}
\end{figure}

We next examine in detail the global suspension viscosity for the emulsions with coalescence.
The effective viscosity for different volume fractions, capillary number and viscosity ratio is shown in figure \ref{fig:effective}, together with the experiments in \cite{Caserta2012}. The rheological curve exhibits a negative curvature, unlike rigid and deformable particles \citep{Guazzelli2018,Rosti2018a}. 
On the left panel of the figure, we display the results for emulsions with viscosity ratio equal to one: the effective viscosity is always greater than the outer fluid viscosity and exhibit a maximum approximatively around $\phi = 0.2$. As Ca decreases, droplets become less deformable leading to an increase of the effective viscosity whereas, for the higher capillary case, the decrease of effective viscosity at high volume fraction is more prominent. When the disperse phase viscosity decreases (right panel of the same figure \ref{fig:effective}) the peak moves to lower volume fractions, about $\phi = 0.1$; in addition, the effective viscosity of the emulsion becomes smaller 1 at larger volume fraction, when increasing the volume of the low viscosity fluid. To check the effect of the droplet to system size ratio, we have also performed simulations with droplets with half the initial radius, shown in black in figure \ref{fig:effective}(right panel):  also in this case the curvature of the effective viscosity is negative. The differences in the values of the effective viscosity are a consequence of the different capillary number in the case of smaller droplets with same interfacial tension.

\cite{Caserta2012} performed experiments of emulsions in shear flow and reported a phase separation associated to the negative curvature of the curve of the effective viscosity versus the volume fraction. This separation results in bands that are aligned in the direction of the flow and alternated in the vorticity direction; the process of banding has been observed only for viscosity ratio smaller than 1. We can observe that in the case of $\lambda = 1$ the concavity of the viscosity curve is only marginally negative, while for $\lambda < 1$ the sign of the curvature is clearly negative both experimentally and numerically. For this reason banding phenomenon was visible only in emulsions with $\lambda < 1$. Because the characteristic width of the bands is of the order of the gap between the walls we cannot reproduce with the present simulations this separation of phase. For this reason we compare results only for the case with $\lambda = 1$, when the distribution of the dispersed phase in homogeneous in space.
The numerical results in figure \ref{fig:effective} are qualitatively in good agreement with the experimental results, also reported in the same figure. It is difficult to perform an exact comparison between the numerical simulations and the experimental results due to the uncertainty on the initial size of the droplets. The initial size distribution in the experiment is a consequence of the application of a strong pre-shear, higher than the one corresponding to the critical capillary number for breakup, to fragment the droplets and remove any effect of the initial configuration. By knowing the value of the pre-shear ($40$ s\textsuperscript{-1}) and assuming that the droplets have a monodisperse radius distribution corresponding to the critical capillary number of 0.5 (for $\lambda = 1$), we can estimate the initial value of the radius. With this estimated radius the $Ca$ for the data in figure \ref{fig:effective} corresponding to the experiments is 0.12. Therefore, the small mismatch between our initial $Ca$ and the estimate of the experiment can be most likely explained by the coalescence efficiency, which is close to unity in our simulations.

It is worth mentioning that, shear banding is a phase separation process observed only for viscosity ratio smaller than unity and in large domains, in the vorticity direction, and over a very long time, \emph{i.e.} more than 1000 shear units. In this case the dispersed phase distribution is not homogenous and there is a reduction in the effective viscosity of the system \citep{Caserta2012}. It is common in the literature to distinguish between the \emph{constitutive curve}, defined as the effective viscosity as function of the volume fraction for an homogeneous microstructure and shear rate, and the \emph{flow curve}, effective viscosity vs volume fraction measured in the rheometer which can include vorticity banding. In this work we force the dispersed phase to be approximatively homogenous due to lateral confinement, hence the curve reported here should be considered as the \emph{constitutive curve} of the emulsion. This should be compared to experimental measures run at short time (see \cite{Caserta2012}) when the bands have not yet formed. The effect of the vorticity banding on the effective viscosity will be object of future works.

\subsection{Rheology of repulsive emulsions}

\begin{figure}
  \centering
  \includegraphics[width=0.7\textwidth]{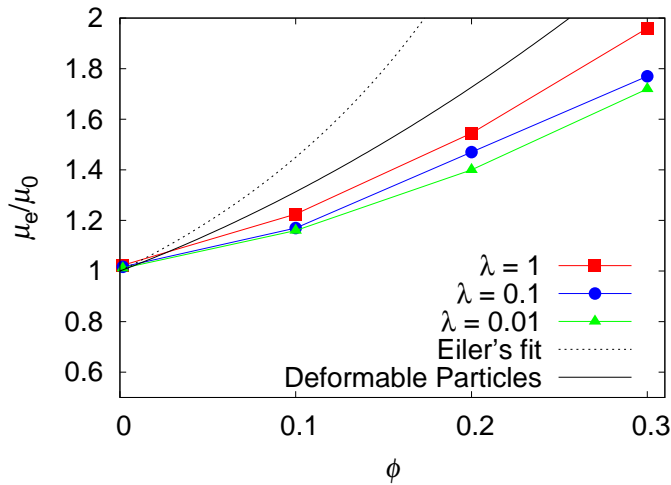}
  \caption{Effective viscosity $\mu_e$ for the cases with collision force and $Ca = 0.1$. The line denoted as "deformable particles" corresponds to the fit proposed in \cite{Rosti2018a} for initially spherical viscoelastic particles with same capillary  number based on the material shear elastic module.}
  \label{fig:effective-collision}
\end{figure}

In this section we study the emulsion behavior when suppressing the coalescence. This is equivalent to assuming that the characteristic drainage time tends to infinity; hence, the collision dynamics is faster than the coalescence time scale and the droplets never merge. We have therefore performed simulations for the same parameters as in the results discussed in the previous section, now
with the collision force given by \eqref{eqn:force}. The effective shear viscosity obtained when varying the viscosity ratio and the volume fraction $\phi$
are reported in figure \ref{fig:effective-collision}.

When the coalescence is prohibited, the curve viscosity vs concentration exhibits a positive curvature, as for the case of rigid and deformable particles. In the same figure, we display also the Eilers fit, valid for solid particles, and the fit from \cite{Rosti2018a} for deformable particles. In particular, \cite{Rosti2018} have shown that it is possible to 
estimate the effective viscosity of a suspension of deformable particles with the Eilers formula by computing an effective volume fraction based on the mean deformation of the particles (see appendix \ref{appB} for more details). The results in figure \ref{fig:effective-collision} demonstrate that, in the absence of coalescence, emulsions behaves as suspensions of deformable particles. 

The change of sign in the curvature can be explained by examining the stress components, as shown previously for droplets coalescing. In this case an additional force, the collision force, needs to be included in the stress budget, which is treated in the same way as the interfacial tension force:
\begin{equation}
  \label{eqn:stress-balance-collision}
  <2 \mu D_{13}>|_{0} =  - <<\rho uw>> + <<2 \mu D_{13}>> + <<\mathcal{G}>> + <<\mathcal{C}>>
\end{equation}
with $\mathcal{C} = \int_z <F_c>\,dz$ the wall-normal integral of the streamwise component of the collision force.
\begin{figure}
  \centering
  \begin{subfigure}{0.45\textwidth}
    \includegraphics[width=\textwidth]{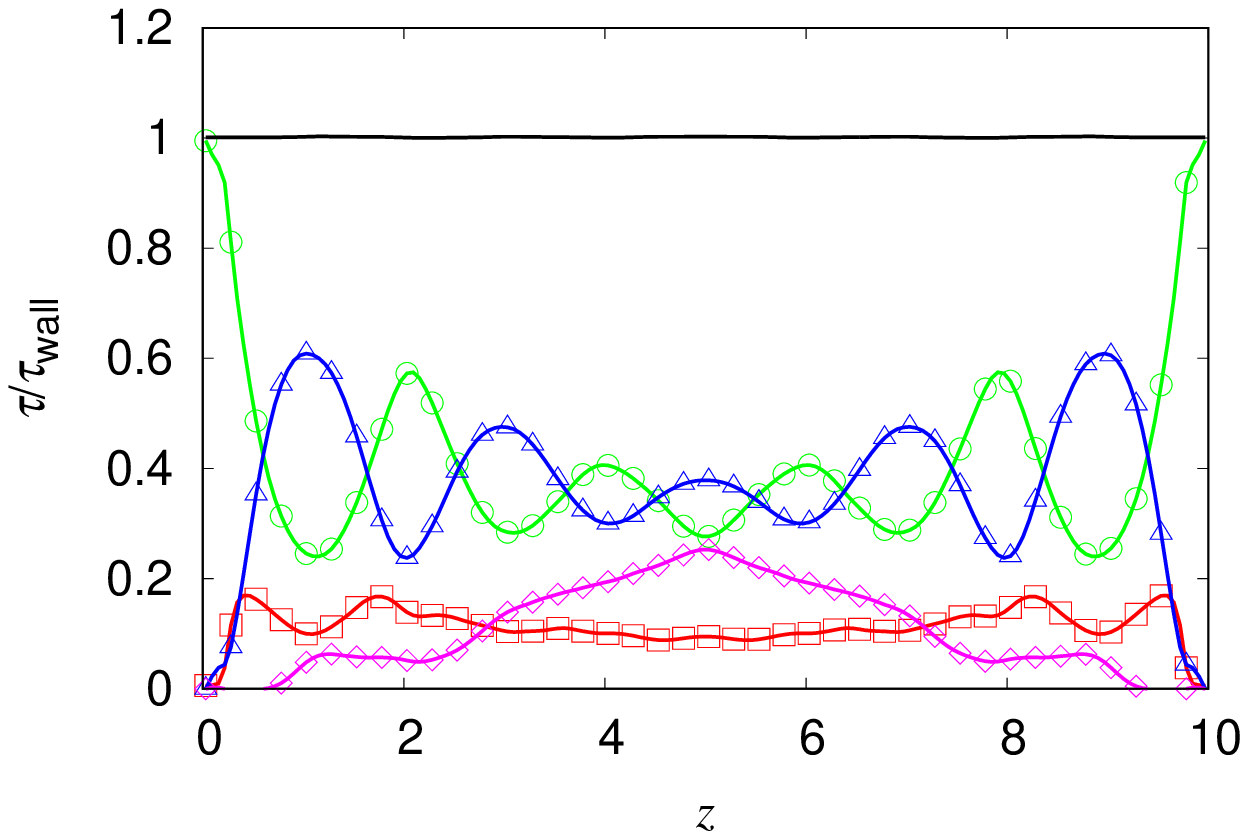}
  \end{subfigure}
  \begin{subfigure}{0.45\textwidth}
    \includegraphics[width=\textwidth]{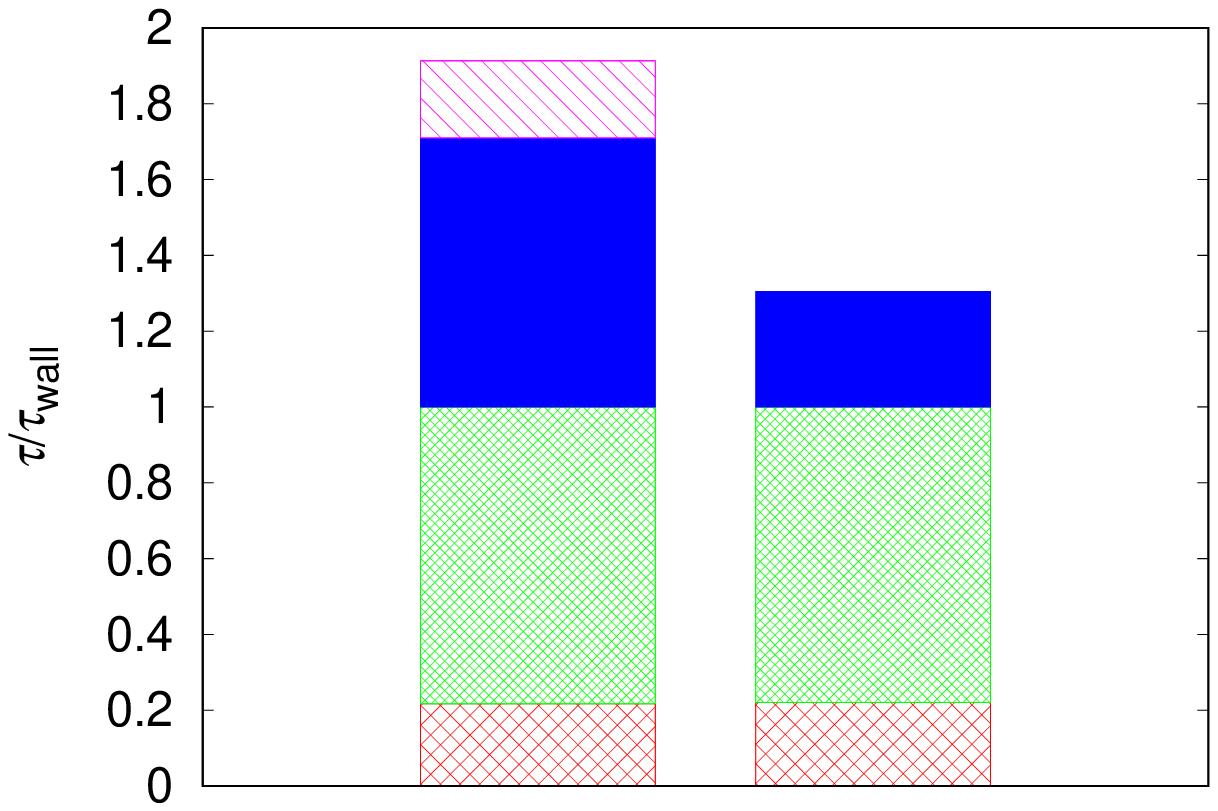}
  \end{subfigure}
  \caption{(Left) Wall-normal distribution ($z$) of the stresses for the case with $Ca = 0.1$, $\phi = 0.3$, $\lambda = 1.0$ and collision force: disperse phase viscous stress (\textcolor{red}{$\Box$}); carrier fluid viscous stress (\textcolor{green}{$\circ$}); interfacial force (\textcolor{blue}{$\triangle$}); collision force (\textcolor{magenta}{$\diamond$}). The stresses are normalized with the wall shear stress. (Right panel) Histogram of the stress components for the case with $Ca = 0.1$, $\phi = 0.3$ with (left column) and without (right column) the collision force: interfacial force (solid blue); carrier fluid viscous stress (green dense net); disperse phase viscous stress (red sparse net); collision force (purple oblique bars). The stresses are normalized with the single-phase wall shear stress.\label{fig:stresses-collision}}
\end{figure}
\begin{figure}
  \centering
  \begin{subfigure}{0.45\textwidth}
    \includegraphics[width=\textwidth]{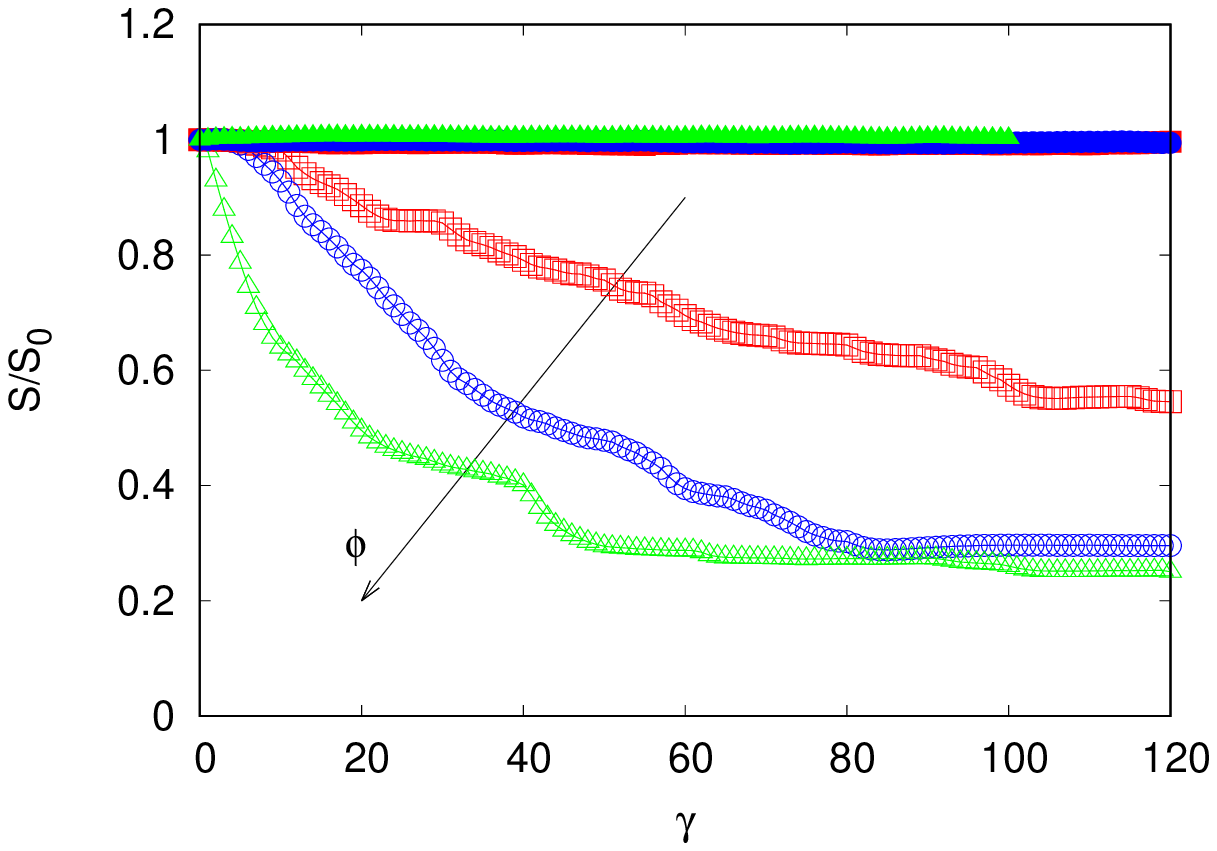}
  \end{subfigure}
  \begin{subfigure}{0.45\textwidth}
    \includegraphics[width=\textwidth]{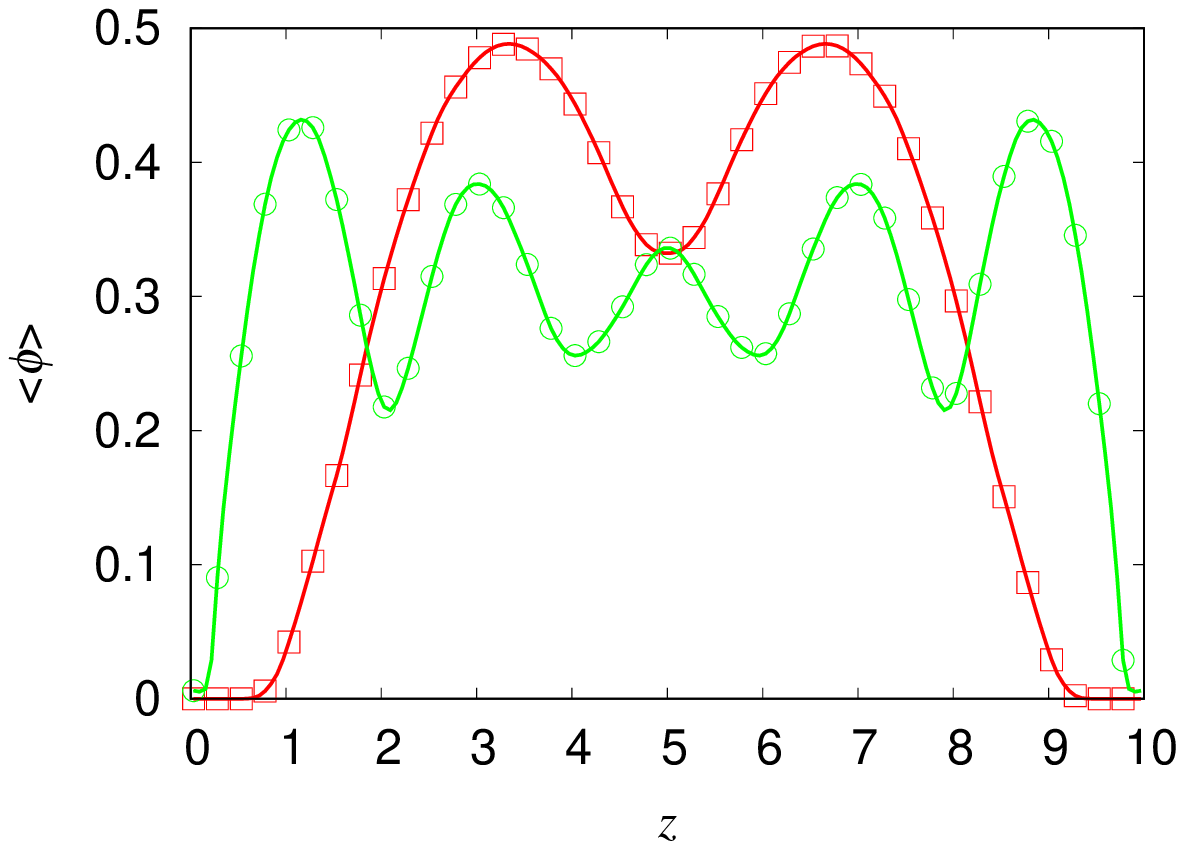}
  \end{subfigure}
  \caption{(Left panel) Time history of the total surface for different volume fraction and $\lambda = 0.01$. Cases without collision force: $\phi = 0.1$ (\textcolor{red}{$\Box$}), $\phi = 0.2$ (\textcolor{blue}{$\circ$}), $\phi = 0.3$ (\textcolor{green}{$\triangle$}); cases with collision force: same color and solid symbols. (Right panel) Comparison of the wall normal distribution of the average volume fraction $<\phi>$ for the cases with $\phi = 0.3$ and $Ca = 0.1$: without collision force (\textcolor{red}{$\Box$}); with collision force (\textcolor{green}{$\circ$}).\label{fig:surface}}
\end{figure}

The stress budget for the case with $Ca = 0.1$, $\phi = 0.3$ and $\lambda = 1$ is shown in figure \ref{fig:stresses-collision}(left panel). The interfacial tension contribution is now of the same order of magnitude of the outer fluid viscous stress, whereas in absence of collision force it is about 2 times smaller, as shown in the right panel. By comparing the stress budget for the case with collision force and without collision force (figure \ref{fig:stresses-collision} right), we notice that the increase in effective viscosity is mostly due to the interface tension term, with a small contribution due to the collision force, which is about 10\% of the total stress. 
Without the collision force, coalescence decreases the total interface area which leads to a reduction of the energy associated to the surface tension. 
This is clearly demonstrated by computing the time history of the total surface. 
In figure \ref{fig:surface} (left panel) the total surface, normalized with the initial value, is displayed for different volume fractions at viscosity ratio $\lambda = 0.01$ for the simulations with and without the collision force. The latter cases exhibit a strong decrease of the total surface before reaching a regime configuration, associated with  a reduction of the interfacial area up to the 80\% of the initial value for the highest volume fraction considered. This statistically steady state is reached faster for the case with higher volume fractions. On the contrary, when coalescence is prohibited the total surface area slightly increases, owing to the deformation of the individual droplets. This explains the increased surface tension contribution to the stress budget observed in the absence of coalescence.

\begin{figure}
  \centering
  \begin{subfigure}{0.45\textwidth}
    \includegraphics[width=\textwidth]{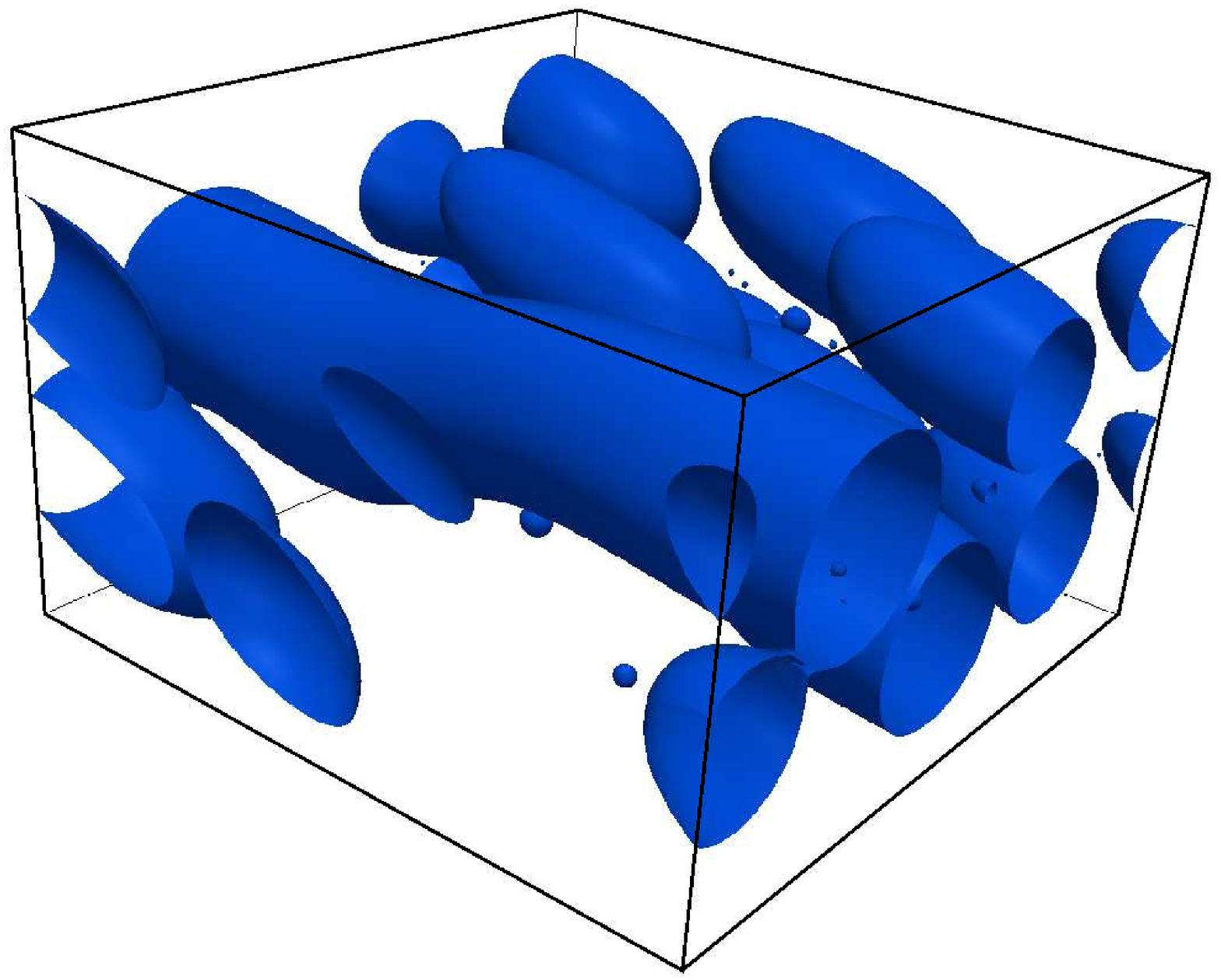}
  \end{subfigure}
  \begin{subfigure}{0.45\textwidth}
    \includegraphics[width=\textwidth]{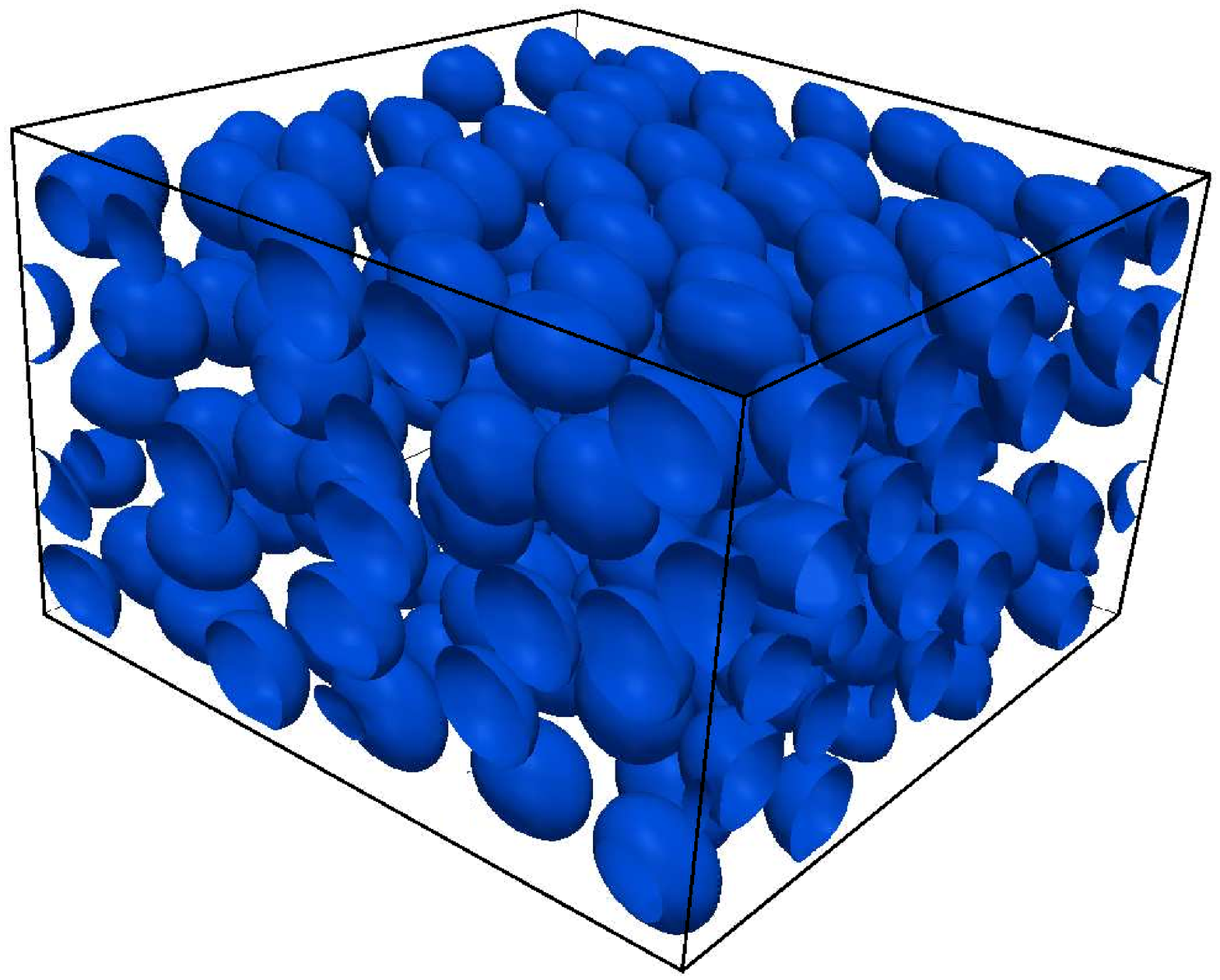}
  \end{subfigure}
  \caption{Instantaneous droplet distribution for emulsions with volume fraction $\phi = 0.3$ and capillary number $Ca = 0.1$ extracted from the simulation without collision force (left) and with collision force (right).\label{fig:visual}}
\end{figure}

Figure \ref{fig:surface}(right) depicts the wall-normal distribution of the average (in the homogeneous $x$ and $y$ directions ) local volume fraction $<\phi>$ for the cases with $\phi = 0.3$ and $Ca = 0.1$. 
Here, we note a significant increase of the volume fraction close to the walls for the cases with collision force, which can be associated to the increased effective viscosity. On the other hand, in the cases with coalescence the larger droplets are localized more towards the center of the domain. This result is in agreement with previous experimental observations \citep{Hudson2003,Caserta2005} where the migration was attributed to the combined action of wall migration and shear-induced diffusion of drops. This is also confirmed in figure \ref{fig:visual} where we report a snapshot of the droplet distribution for the cases in figure \ref{fig:surface}(right).

\begin{figure}
  \centering
  \begin{subfigure}{0.45\textwidth}
    \includegraphics[width=\textwidth]{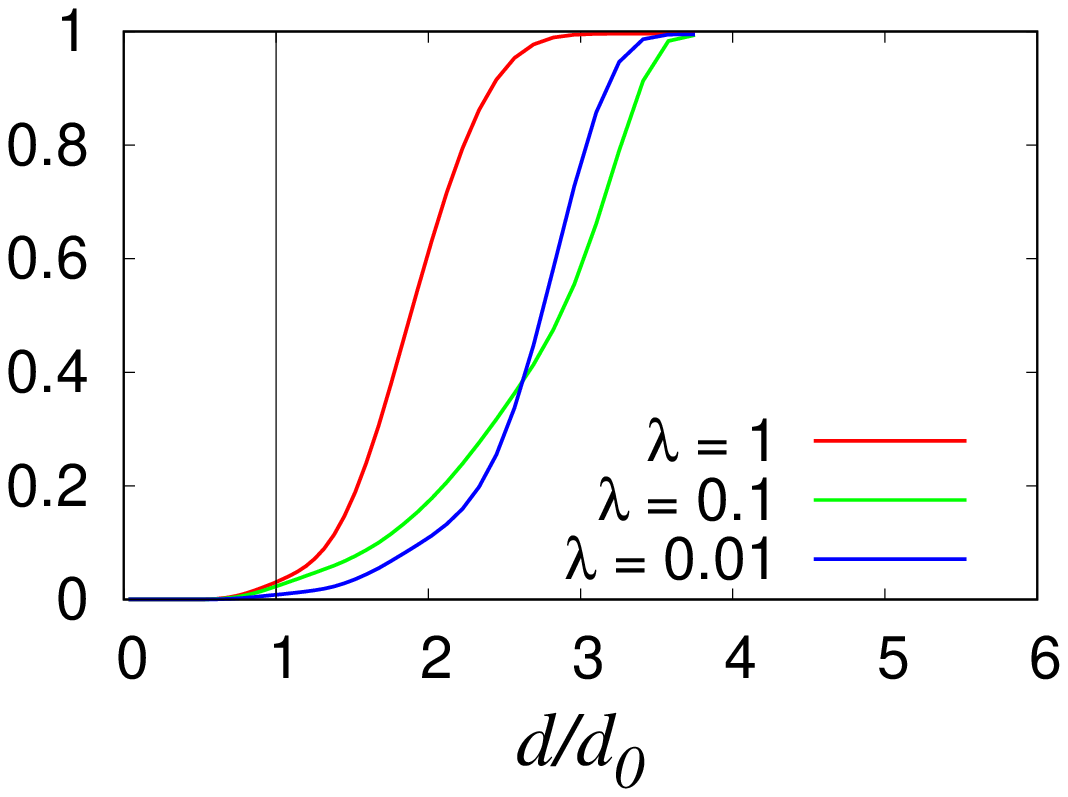}
  \end{subfigure}
  \begin{subfigure}{0.45\textwidth}
    \includegraphics[width=\textwidth]{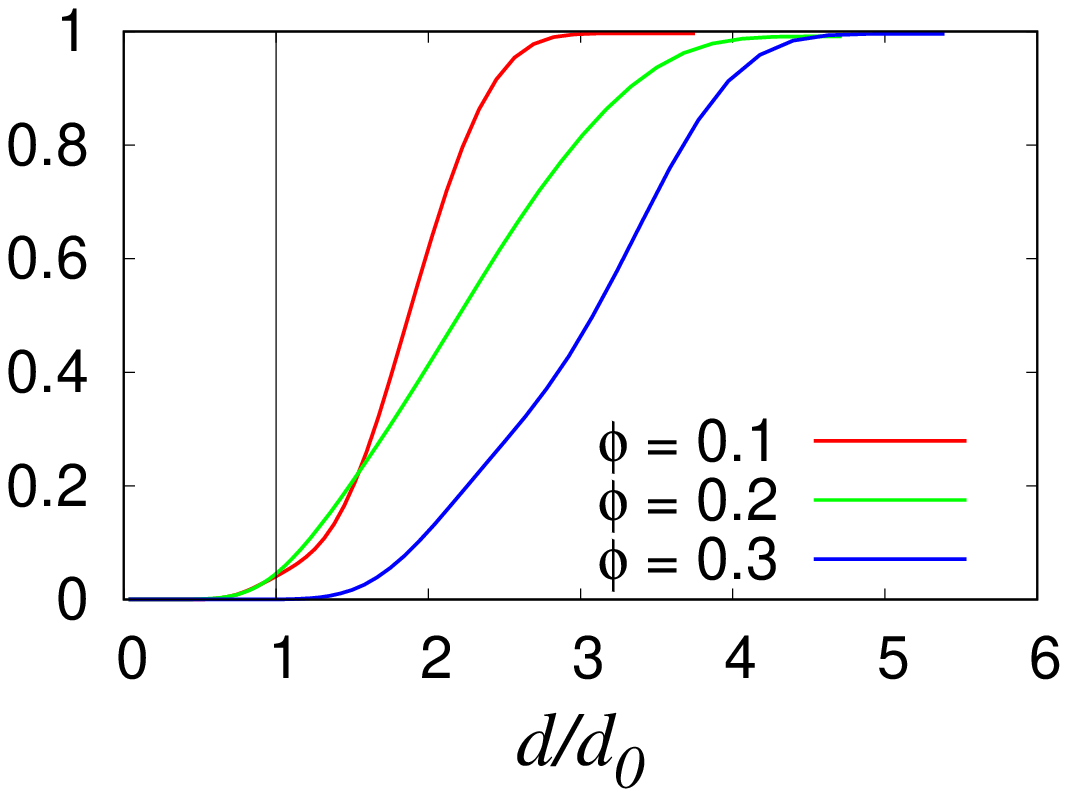}
  \end{subfigure}
  \caption{Cumulative volume distribution as a function of the equivalent diameter, normalized with the initial value, for the case with $\phi = 0.1$ and different $\lambda$ (left panel) and for the cases with different volume fraction and $\lambda = 1$. The black line represents the initial distribution, which is an Heaviside on the initial value of the diameter.\label{fig:cumulative}}
\end{figure}

A measure of the coalescence is the droplet size distribution in the system. To study this, we evaluate the volume distribution in the whole domain at steady state and draw the cumulative distribution in figure \ref{fig:cumulative} for the cases without collision force. On the horizontal axis we report the value of the equivalent diameter computed with the volume of each dispersed droplet, 
normalized with the initial value, and on the vertical axis the fraction of the total volume occupied by droplets of size smaller and equal to the corresponding abscissa. At volume fraction $\phi = 0.1$ the maximum value of the diameter is about 3.5 the initial one and, by reducing the viscosity ratio the distribution shows a reduction of droplets with small diameter due to the increase of coalescence. By increasing the volume fraction droplets merge into larger structures which have a maximum equivalent diameter of about 5 times the initial one. In the simulations with collision force, the mean spherical equivalent diameter is the same for all the droplets and equal to the initial value, depicted by the black line in figure \ref{fig:cumulative}.

\subsection{Flow topology and normal stress}

\begin{figure}
  \centering
  \begin{subfigure}{0.49\textwidth}
    \includegraphics[width=\textwidth]{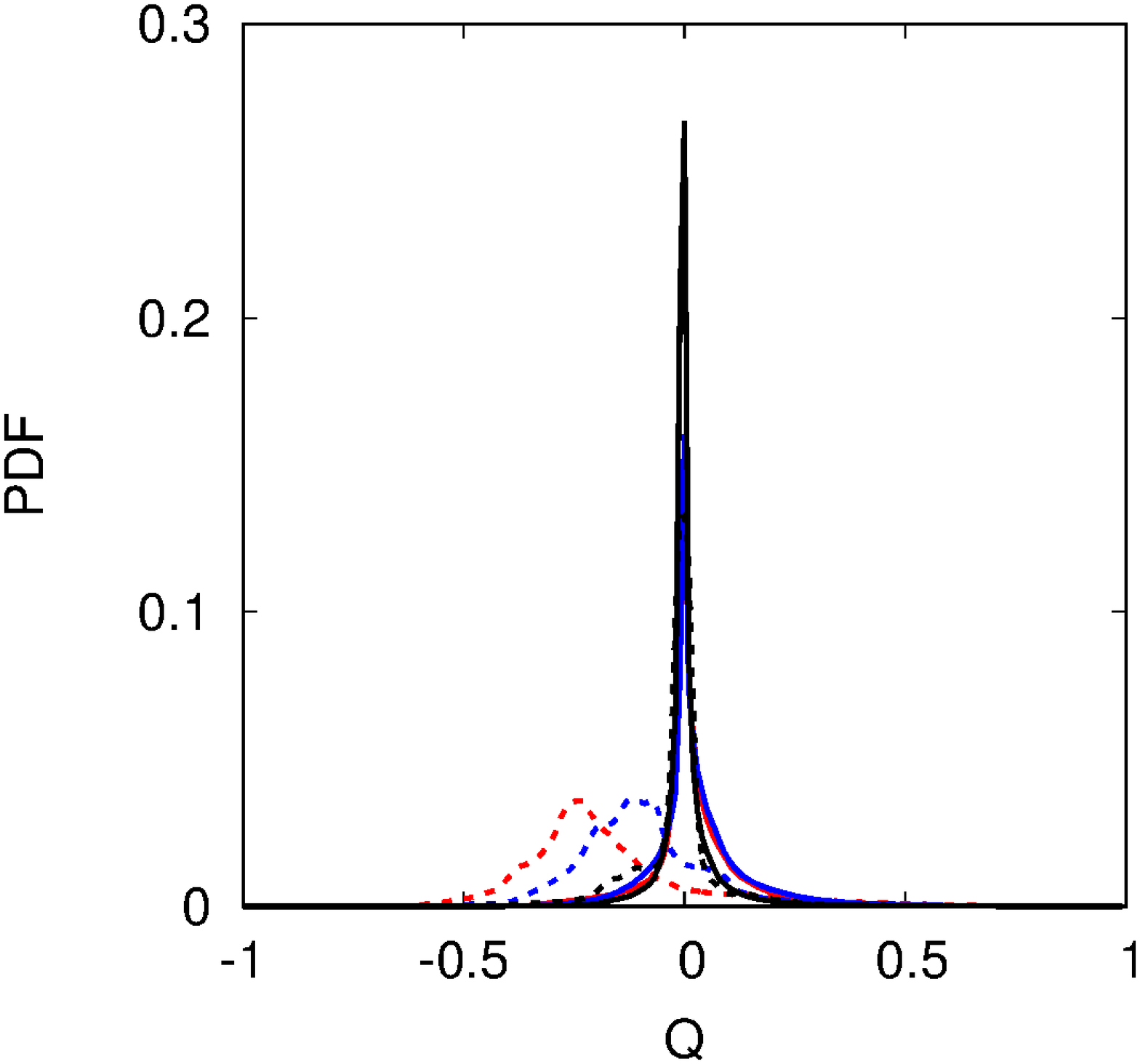}
  \end{subfigure}
  \begin{subfigure}{0.49\textwidth}
    \includegraphics[width=\textwidth]{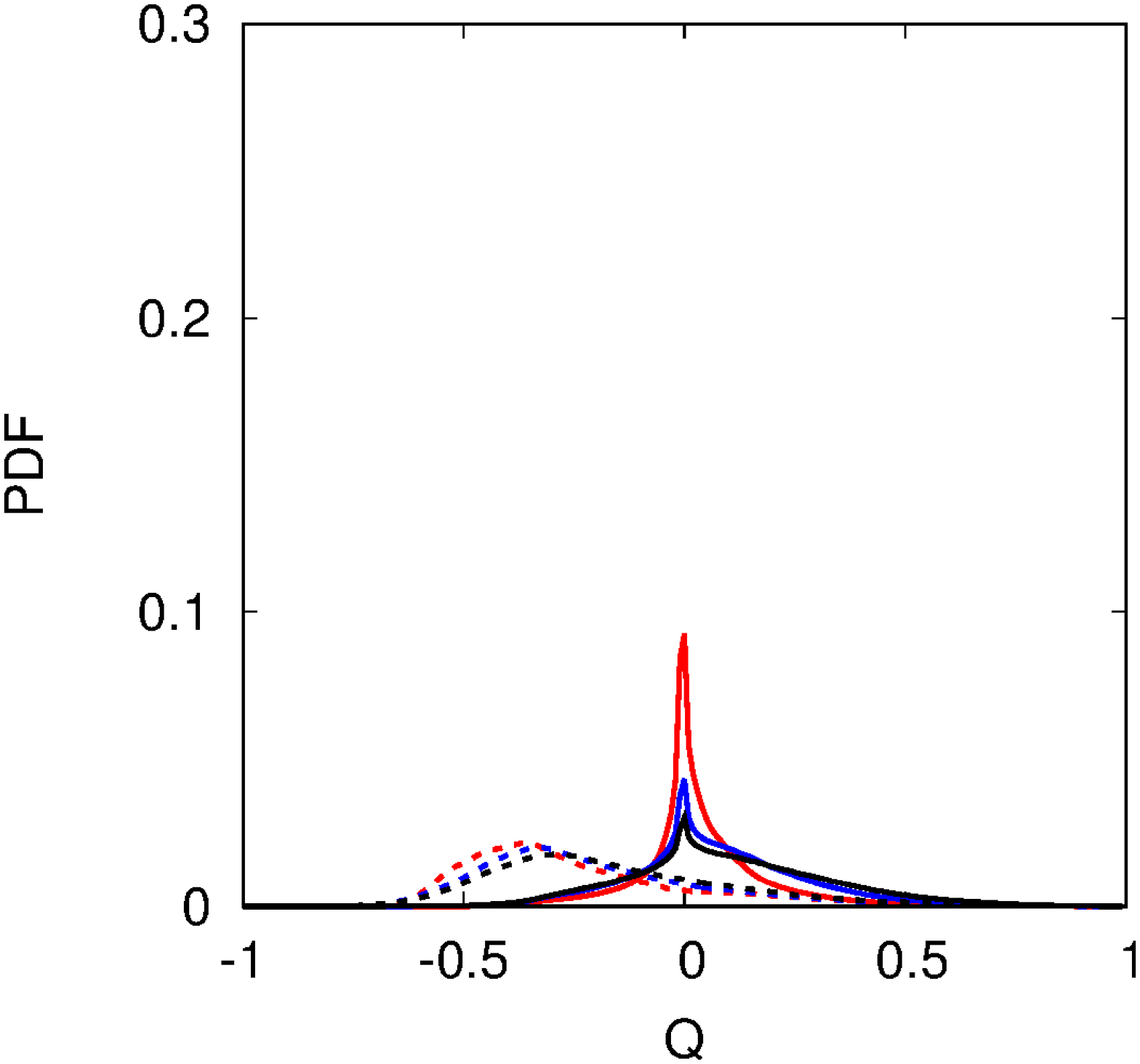}
  \end{subfigure}
  \caption{PDF of the flow topology parameter for simulations without repulsive force among droplets (left) and with (right): solid lines represent the outer fluid domain and dashed lines the dispersed phase. Colors refer to different volume fractions: $\phi = 0.1$ red, $\phi = 0.2$ blue, $\phi = 0.3$ black. \label{fig:Q}}
\end{figure}

Normal stresses can arise in emulsions when sheared and provide evidence for the non-Newtonian behaviour of the system \citep{Loewenberg1996,Pal2011,Srivastava2016}. To evaluate the effect of the coalescence and of the volume fraction on the extensibility of the flow we compute the flow topology parameter \citep{Vita2018,Rosti2018b} defined as
\begin{equation}
  Q = \frac{D^2-\Omega^2}{D^2+\Omega^2}
\end{equation}
where $D^2=D_{ij}D_{ij}$ and $\Omega^2=\Omega_{ij}\Omega_{ij}$ with $\mathbf{\Omega} = (\nabla \mathbf{u}^T - \nabla\mathbf{u})/2$ the rate of rotation tensor. When $Q = -1$ the flow is purely rotational, whereas regions with $Q = 0$ represent pure shear flow and those with $Q = 1$ elongational flow. In figure \ref{fig:Q} we display the 
probability distribution function (PDF) of Q for different volume fractions and for simulations with and without collision force. We observe that in presence of coalescence the outer fluid has a peak of the PDF in correspondence of $Q = 0$, which implies that the flow is almost pure shear flow as in the case of single phase Couette flow. The dispersed phase, instead, shows a pick for negative values of Q (rotational flow), which moves towards zero increasing the volume fraction. When the collision force is applied, we note a strong reduction of the peaks in the outer fluid distribution, compensated by an increase of the underlying area of the curve for positive values of Q. This entails that when the coalescence is prohibited we find in the external flow regions not only of shear flow but also of extensional flow, which is related to the velocities in the gaps between the droplets. For these cases we also observe that the dynamics of the dispersed phase flow is substantially unaffected by the volume fraction and has always components of rotational flow (inside the deforming droplets).

We conclude our analysis by computing the normal stress difference following the Batchelor's formulation \citep{Batchelor1970,Srivastava2016}. Here, however,  we cannot follow the same methodology used for the shear stress, because it is not possible to obtain the integration constant 
needed to determine the function $\mathcal{G}$
when integrating in the homogeneous directions. Following Batchelor's formulation, the bulk stress $<<\bm{\tau}>>$ can be decomposed into the sum of the stress due to the outer fluid, in the case of zero volume fraction, $<<\bm{\tau}^0>>$ and that arising from the presence of the disperse phase $<<\bm{\tau}^1>>$
\begin{equation}
  <<\bm{\tau}>> = -<<p>>\mathbf{I} + <<\bm{\tau}^0>> + <<\bm{\tau}^1>>
\end{equation}
The last term on the RHS can be further decomposed as the sum of three terms: the first due to the viscosity difference $<<\bm{\tau}^{\mu}>>$, the second to the interfacial tension $<<\bm{\tau}^{\sigma}>>$ and the third to the perturbation in the velocity field induced by the presence of the droplets $<<\bm{\tau}^{ptb}>>$
\begin{equation}
  \begin{aligned}
    <<\bm{\tau}^1>> = <<\bm{\tau}^{\mu}>> + <<\bm{\tau}^{\sigma}>> + <<\bm{\tau}^{ptb}>> = \\
    \frac{\mu_1 - \mu_0}{\mathcal{V}} \int_{\mathcal{S}} \left(\mathbf{u}\mathbf{n} + \mathbf{n}\mathbf{u}\right)\,d\mathcal{S} - \frac{\sigma}{\mathcal{V}}\int_{\mathcal{S}}\left(\mathbf{n}\mathbf{n} - \frac{\mathbf{I}}{3}\right)\,d\mathcal{S} - \frac{1}{\mathcal{V}} \int_{\mathcal{V}} \left(\rho\mathbf{u}'\mathbf{u}'\right)\,d\mathcal{V}.
  \end{aligned}
\end{equation}
where $\mathcal{S}$ is the total surface of the droplets and $\mathcal{V}$ is the averaging volume, \emph{i.e.} the computational domain. The normalized normal stress difference are then defined as
\begin{equation}
  \begin{aligned}
    N_1 = \frac{<<\tau^1_{xx}>>-<<\tau^1_{zz}>>}{\mu^0 \dot{\gamma}} \\
    N_2 = \frac{<<\tau^1_{zz}>>-<<\tau^1_{yy}>>}{\mu^0 \dot{\gamma}}.
  \end{aligned}
\end{equation}
When $N_1$ is greater then zero and $\tau^1_{xx} > \tau^1_{zz}$ droplets are elongated in the direction of the flow and compressed in the wall-normal direction. The opposite configuration corresponds to negative values of $N_1$. 

\begin{figure}
  \centering
  \begin{subfigure}{0.49\textwidth}
    \includegraphics[width=\textwidth]{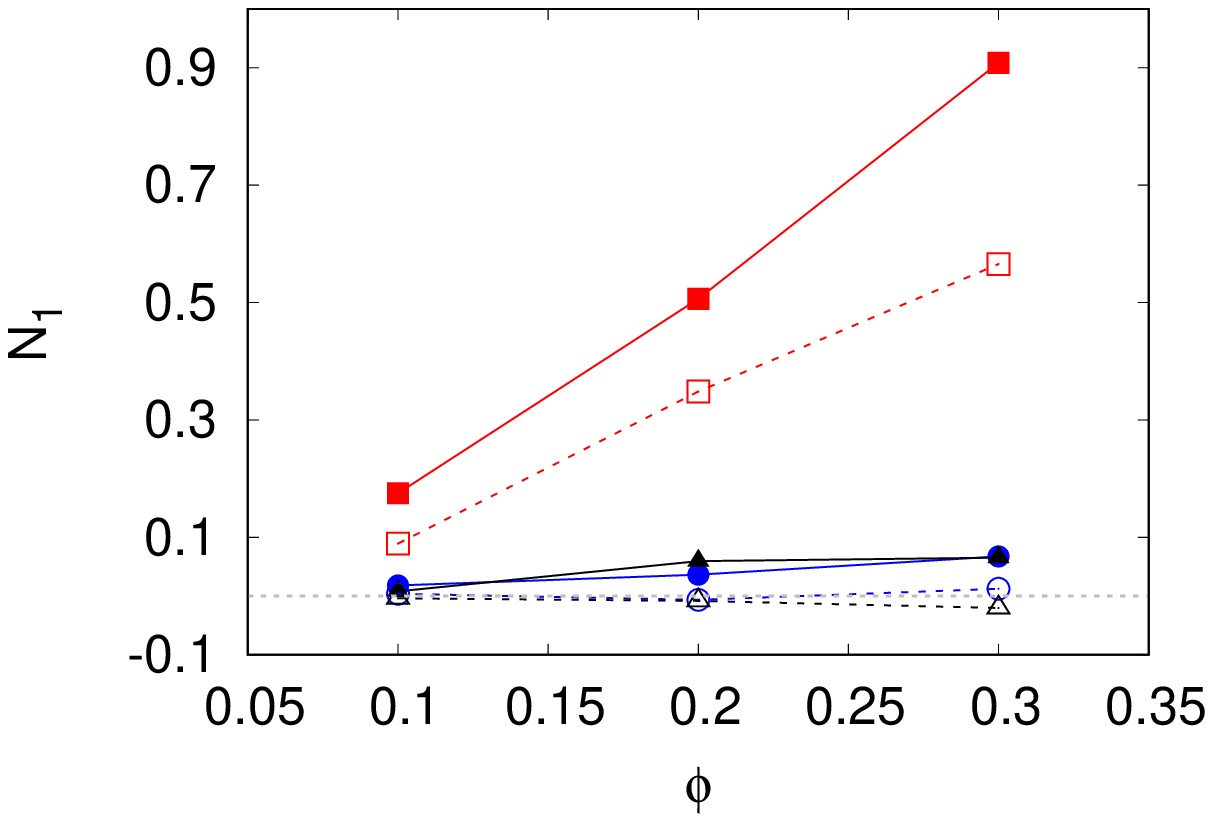}
  \end{subfigure}
  \begin{subfigure}{0.49\textwidth}
    \includegraphics[width=\textwidth]{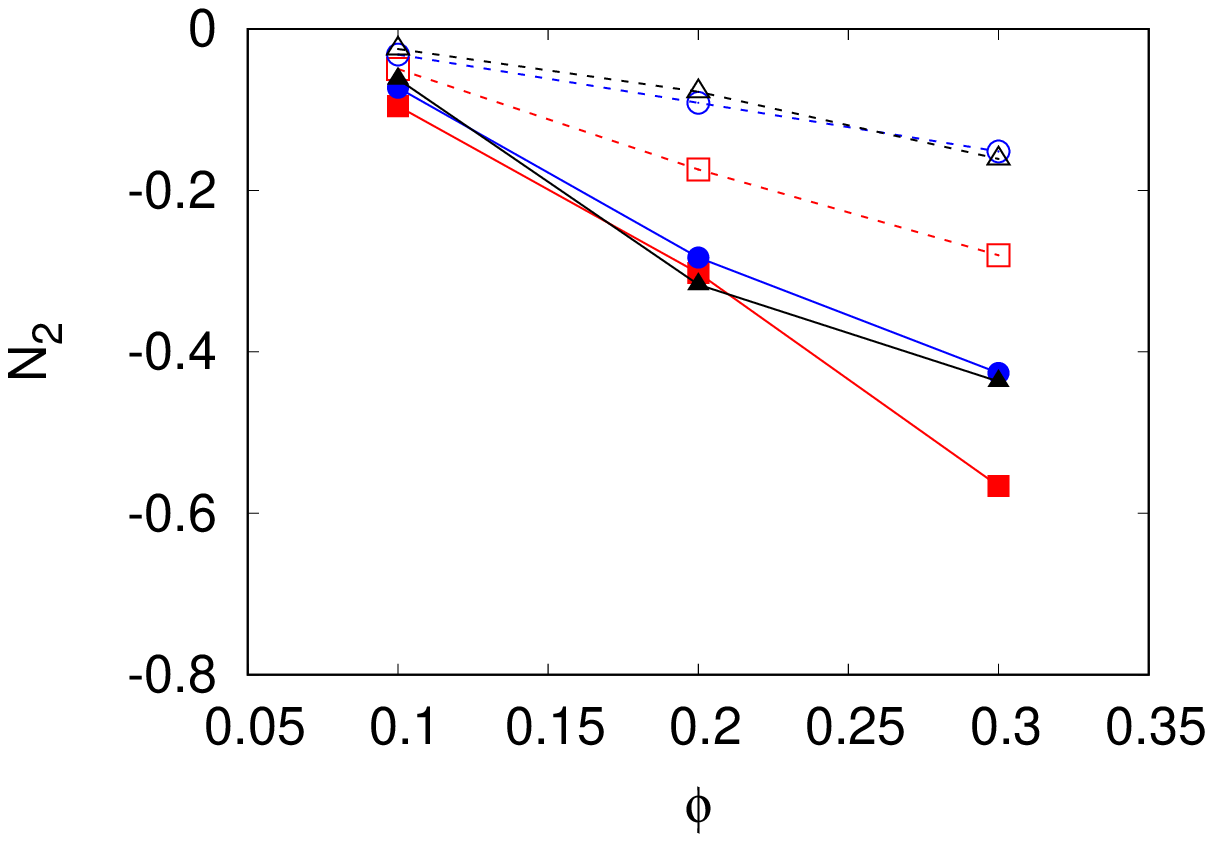}
  \end{subfigure}
  \caption{Normal stress differences for the cases with repulsive force among the droplets (open symbols and dashed lines) and without (solid symbol and solid lines).  Colors represent the different viscosity ratios under investigation: $\lambda = 1$ red, $\lambda = 0.1$ blue, $\lambda = 0.01$ black.\label{fig:normalstress}}
\end{figure}

In figure \ref{fig:normalstress}, we report the values of $N_1$ and $N_2$ computed from our simulations. 
First, we observe that the magnitude of the normal stress difference is larger for the cases without collision force. This can be explained noting that when the collision force is 
turned on the droplets exhibit smaller deformation due to the higher packing inside the domain (see figure \ref{fig:visual}). The magnitude of the normal stress differences decreases with the viscosity ratio and increases with the volume fraction. The data in the figure also reveal that the second normal difference is always negative whereas the first normal difference is positive for $\lambda = 1$ and $\lambda = 0.1$ and becomes negative for smaller viscosity ratio. Additionally, the ratio of $N_1$ and $N_2$ is almost constant with the volume fraction, showing that they have the same dependency on $\phi$.

\section{Conclusion}\label{sec:conclusion}

We have studied the rheology of emulsions under shear flow in the dilute and moderate concentration regime by means of numerical simulations. The multiphase flow dynamics is governed by the Navier-Stokes equations while the interface is tracked in time employing a Volume of Fluid technique. This approach naturally allows us to take into account the coalescence of droplets, which significantly affects the rheological behaviour of emulsions. To single out the effect of coalescence, we have therefore developed an Eulerian collision model which allows to delay or fully prohibit the merging of droplets. We have reported here simulations at different capillary number, viscosity ratio and volume fraction with a constant Reynolds number, small enough to assume inertial effects to be negligible. In this work we have focused on the characterisation of the constitutive curves of emulsions by simulating approximatively homogeneous distributions of the dispersed phase and neglecting the vorticity banding observed in experiments at viscosity ratio smaller than unity.

We show that the curvature of the rheological curve (effective viscosity versus volume fraction) is negative when coalescence is allowed whereas it changes sign when we introduce the collision force which prevents the merging. The decrease of the effective viscosity in the former case is a consequence of the reduction of the total surface of the system, which in turn reduces the contribution to the stress tensor due to the interface tension stress. When the coalescence is prohibited, this term is responsible for about half of the total effective viscosity and emulsions behave similarly to suspensions of deformable particles, as further demonstrated in the Appendix. 

In the case of coalescence, we observe the formation of relatively large droplets, migrating towards the channel centre. Moreover, by examining the probability distribution function of the flow topology parameter, we show that the external flow is mostly shear while the dispersed phase exhibits some rotational flow regions, which become smaller when increasing the volume fraction (on average large droplets). On the other side, when we introduce the collision force, the dominant presence of shear flow in the outer fluid decreases and regions of extensional flows emerge. The dispersed phase, in this case characterised by smaller deformed droplets in the flow, is mostly unaffected by the volume fraction and shares, equally distributed, regions of shear and rotational flow. 
Furthermore, analysis of the droplet size distribution at steady state reveals that in the presence of coalescence the mean equivalent diameter of the droplets in the emulsions increases up to three times the initial value and is function of both volume fraction and viscosity ratio. For simulations with collision force, instead, every droplet has an equivalent diameter equal to the initial one.

To characterise the visco-elastic system behavior, we show that the first normal stress difference is positive and the second negative, as in suspensions of capsules and deformable particles. The magnitude of the normal stress difference always increases with the volume fraction; also noteworthy we find an inversion of the first normal difference for small values of the viscosity ratio. 

To conclude, it is worth noticing once more that we have investigated two limiting cases, one with coalescence efficiency tending to unity and one with efficiency approaching zero. In a real scenario, the coalescence efficiency is likely to have intermediate values and the behaviour of emulsions can therefore differ from the limiting cases considered here. Future works may therefore be devoted to handling both coalescence and collisions in order to simulate different types of emulsions and to study the formation of the vorticity banding. 

\appendix
\section{}\label{appA}

\begin{figure}
  \centering
  \begin{subfigure}{0.49\textwidth}
    \includegraphics[width=\textwidth]{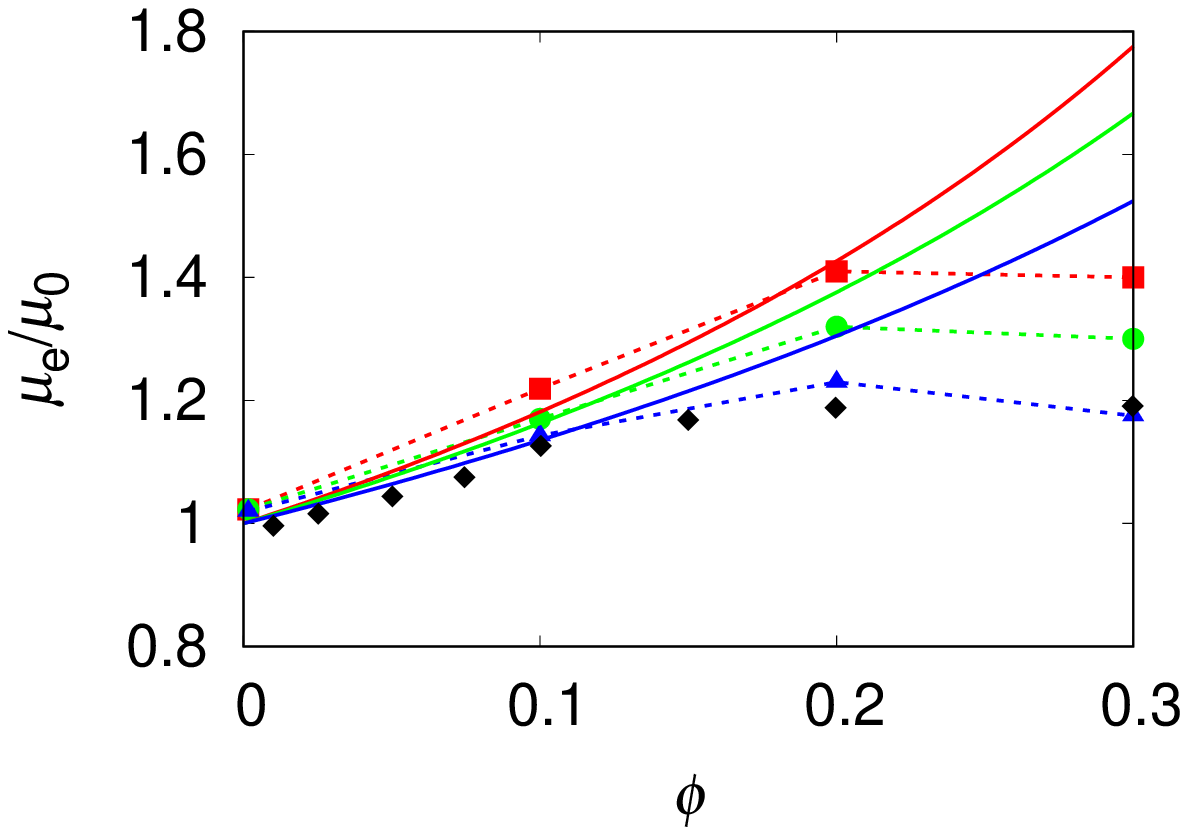}
  \end{subfigure}
  \begin{subfigure}{0.49\textwidth}
    \includegraphics[width=\textwidth]{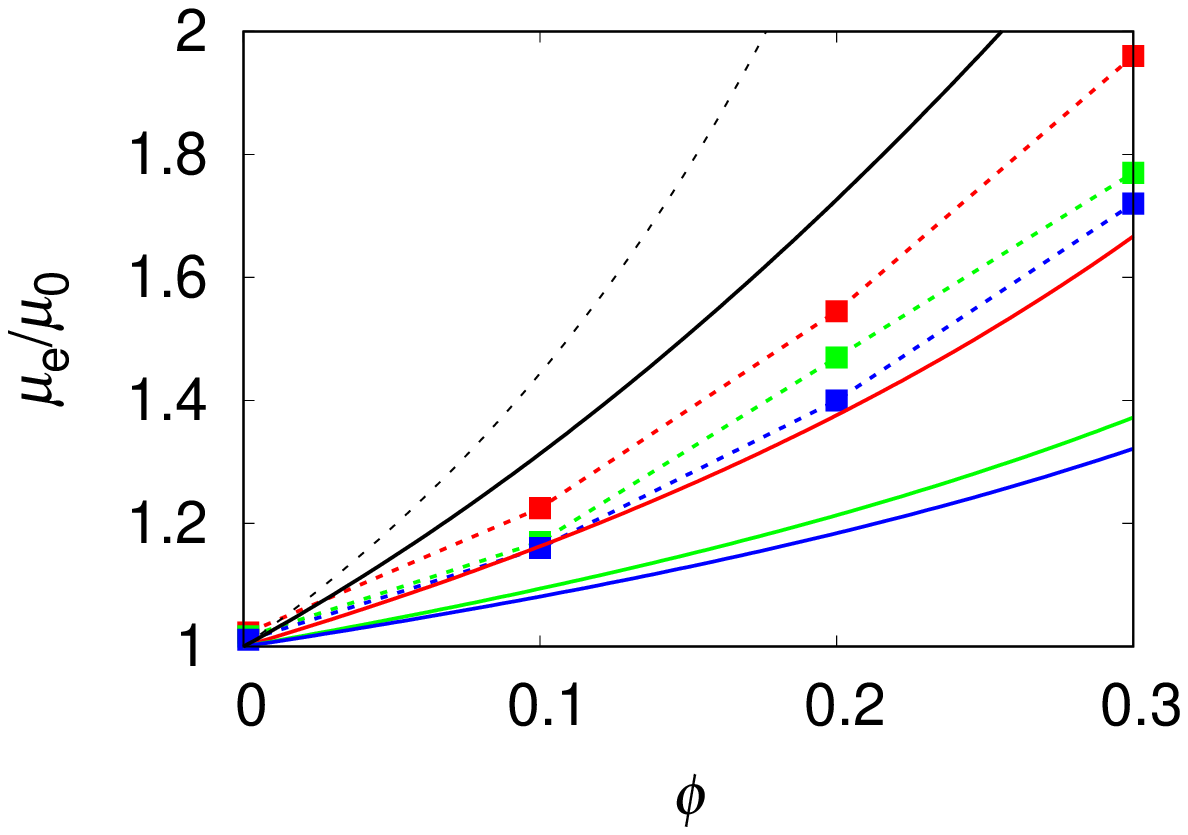}
  \end{subfigure}
  \caption{Effective viscosity versus volume fraction. (Left panel) Simulations with $\lambda = 1$ and different capillary numbers: solid lines represent equation \eqref{eqn:Pal}, dashed lines with symbols represent our numerical simulations. Colors represent different capillary number: $0.05$ red, $0.1$ green, $0.2$ blue. Black diamonds represent experiment by \cite{Caserta2012}. (Right panel) Cases with $Ca = 0.1$ and different $\lambda$: solid colored lines represent equation \eqref{eqn:Pal}, dashed lines with symbols represent our numerical simulations, black solid line represent data from \cite{Rosti2018} and black dashed line represent equation \ref{eqn:Eilers}. Colors represent different $\lambda$: $1$ red, $0.1$ green, $0.01$ blue.   \label{fig:A1}}
\end{figure}

In this appendix, we compare our results with expressions available in literature for suspensions of particles and emulsions. \cite{Pal2003} proposed the following equation for the viscosity of a concentrated emulsion
\begin{equation}
  \label{eqn:Pal}
  \mu_r\left[\frac{M-P+32\mu_r}{M-P+32}\right]^{N-1.25}\left[\frac{M+P-32}{M+P-32\mu_r}\right]^{N+1.25} = \left(1-\frac{\phi}{\phi_m}\right)^{-2.5\phi_m}
\end{equation}
with $\mu_r = \mu_e / \mu_0$, $\phi_m$ the maximum packing volume fraction and
\begin{subequations}
  \begin{align}
    M &= \sqrt{\frac{64}{Ca^2} + 1225\lambda^2 + 1232\frac{K}{Ca}}\\
    P &= \frac{8}{Ca} - 3\lambda \\
    N &= \frac{\dfrac{22}{Ca} + 43.75 \lambda}{M}.
  \end{align}
\end{subequations}
For the comparison reported here, we use $\phi_m = 0.637$, as reported by \cite{Pal2003}. In figure \ref{fig:A1}(left) we display the results of our simulations for $\lambda = 1$ and different capillary numbers alongside equation \eqref{eqn:Pal}, for the same parameters. In the dilute regime, up to 10\% volume fraction, the comparison provides good agreement between simulations and the proposed relation, \eqref{eqn:Pal}, whereas the curves diverge for higher volume fractions. The reason is that equation \eqref{eqn:Pal} cannot reproduce the positive concavity of the effective viscosity versus the volume fraction, as reported in the experiment by \cite{Caserta2012} (also shown in the plot with black diamonds). Additionally, we note that equation \eqref{eqn:Pal} assumes as parameter the maximum packing volume fraction which, in case of coalescence, does not have a clear physical meaning. 

Next, see figure \ref{fig:A1}(right), we compare \eqref{eqn:Pal} with results for the cases with collision force, Eilers formula
\begin{equation}
  \label{eqn:Eilers}
  \mu_r = \left[1 + \frac{1.25\phi}{1-\frac{\phi}{\phi_m}} \right]^2
\end{equation}
and the data from \cite{Rosti2018a} for deformable particles of a viscous hyperelastic material.  Despite the differences in the specific values, the trend is similar for all curves, with \eqref{eqn:Pal} underpredicting the effective viscosity. It is worth noticing that, if the parameter $\phi_m$ is used as a fitting parameter, it is possible to get a better agreement between our results and equation \ref{eqn:Pal}, in particular using $\phi_m \approx 0.4$. 

\section{}\label{appB}
\begin{figure}
  \centering
  \includegraphics[width=0.7\textwidth]{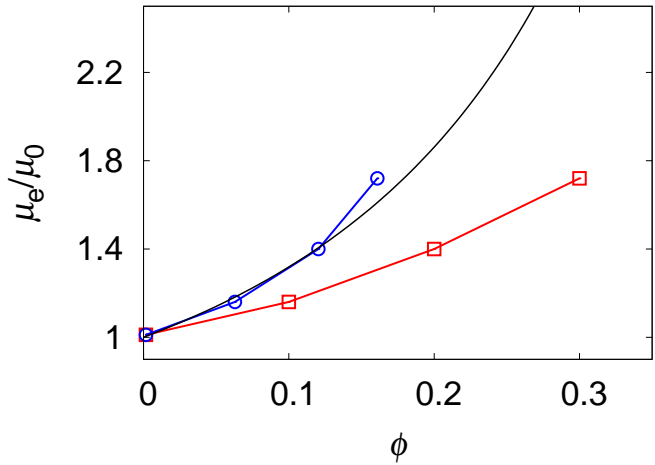}
  \caption{Effective viscosity as a function of the volume fraction $\phi$ (\textcolor{red}{$\Box$}) and of the effective volume fraction $\phi_e$ (\textcolor{blue}{$\circ$}). The black solid line represent the Eilers formula \eqref{eqn:Eilers}. \label{fig:taylor}}
\end{figure}

\cite{Rosti2018a} showed that the effective viscosity of a suspension of deformable particles can be predicted using the Eilers formula \eqref{eqn:Eilers} by modifying the volume fraction to take into account the particle deformation. This can be measured by the Taylor parameter 
\begin{equation}
  \mathcal{T} = \frac{a-b}{a+b}
\end{equation}
where $a$ and $b$ are the semi-major and semi-minor axis of the inscribed ellipse. The authors showed that this fit provides good predictions for different capillary numbers, viscosity ratios, both  for fluid-filled capsules and red blood cells. Here, we wish to show that the same scaling can be applied also for the emulsions considered in this study when coalescence is inhibited. We therefore evaluate the effective viscosity $\phi_e$ based on spheres of radius equal to the semi-minor axis $b$ of the inscribed ellipsoid 
\begin{equation}
  \phi_e = N \frac{\frac{4}{3} \pi b^3}{\mathcal{V}}
\end{equation}
where $N$ is the number of droplets in the computational box of volume $\mathcal{V}$. We then depict in figure \ref{fig:taylor} the effective viscosity of the emulsions as function of the average volume fraction $\phi$ and of the effective volume fraction $\phi_e$. 
The figure shows that indeed the effective viscosity of the emulsion successfully collapses on the Eilers formula, with an error of about 6\% only for the highest volume fraction. It is worth noticing that this scaling works only for the cases with collision force; indeed, in the case of coalescence the shape of the interface is not well approximated by an ellipsoid hence the Taylor parameter is not a good measure of the droplet deformation. Additionally, the curvature of the curve representing the effective viscosity versus the volume fraction is negative and thus it is not possible to collapse the results on the Eilers formula, which is instead monotonic. These results further prove that in the absence of coalesce emulsions behave similarly to suspensions of deformable particles and capsules and that coalescence should be considered to properly describe the rheological response of emulsions.

\section*{Acknowledgements}
The work is supported by the Microflusa project, the European Research Council grant no. ERC-2013-CoG-616186, TRITOS. The Microflusa project receives funding from the European Union Horizon 2020 research and innovation program under Grant Agreement No. 664823. The authors acknowledge computer time provided by SNIC (Swedish National Infrastructure for Computing).

\bibliographystyle{jfm}
\bibliography{biblio}

\end{document}